\documentclass[aps, prd, preprint, onecolumn, tightenlines, notitlepage, superscriptaddress, nofootinbib, preprintnumbers, floatfix,showkeys]{revtex4-1}

\usepackage{amstext}
\usepackage{amssymb}
\usepackage{amsmath,bm}
\usepackage{rotating,graphicx}
\graphicspath{{plots/}}
\usepackage{hyperref}
\usepackage{url}
\usepackage{color}
\usepackage{ulem}
\usepackage[utf8]{inputenc}
\pdfoutput=1
\usepackage[x11names]{xcolor}
\usepackage{textcomp}
\usepackage{booktabs}
\usepackage{epsfig,amsfonts,mathrsfs,amsmath,amssymb,graphicx,color,slashed,multirow}
\usepackage{amsmath,latexsym,amssymb,graphicx,slashed,hyperref,color,enumerate,url,cancel,gensymb}
\usepackage{textcomp}

\newcommand{\be}{\begin{equation}}
	\newcommand{\ee}{\end{equation}}
\newcommand{\bea}{\begin{equation} \begin{aligned}}
	\newcommand{\eea}{\end{aligned} \end{equation}}	
	
\hypersetup{colorlinks,citecolor= nicegreen,linkcolor= nicegreen}
\definecolor{nicered}{rgb}{0.7,0.1,0.1}
\definecolor{nicegreen}{rgb}{0.1,0.5,0.1}
\definecolor{mygreen}{rgb}{0,0.392,0}
\definecolor{mygreen}{rgb}{0,0,0.545}

\newcommand{\orcid}[1]{\href{https://orcid.org/#1}{#1}}

\newcommand {\red} 

\def \GeV{{\mathrm{GeV}}}

\AtBeginDocument{\hypersetup{citecolor=DarkOrange3,linkcolor=Green4,urlcolor=Green4}}
\begin{document}

\title{{\Large A Robust Description of Hadronic Decays in Light Vector Mediator Models}}

\author{Ana Luisa Foguel }\email{afoguel@usp.br}
\thanks{orcid \# \orcid{0000-0002-4130-1200}}
\author{Peter Reimitz}\email{peter@if.usp.br}
\thanks{orcid \# \orcid{0000-0002-4967-8344}}
\author{Renata Zukanovich Funchal}\email{zukanov@if.usp.br} 
\thanks{orcid \# \orcid{0000-0001-6749-0022}}
\affiliation{Departamento de F\'{\i}sica Matem\'atica, Instituto de F\'{\i}sica\\
Universidade de S\~ao Paulo, C. P. 66.318, 05315-970 S\~ao Paulo, Brazil\\[5mm] ~}

\preprint{}

\begin{abstract}

Abelian $U(1)$ gauge group extensions of the Standard Model represent one of the most minimal approaches to solve some of the most urgent particle physics questions and provide a rich phenomenology in various experimental searches.
In this work, we focus on baryophilic vector mediator models in the MeV-to-GeV mass range and, in particular, present, for the first time,
gauge vector field decays into almost arbitrary hadronic final states. Using only very little theoretical approximations, we rigorously follow the
vector meson dominance  theory in our calculations. We study the effect on the total and partial decay widths, the branching ratios, and not least on the present (future) experimental limits (reach) on (for) the mass and couplings of light vector particles in different models. We compare our results to current results in the literature. Our calculations are publicly available in a python package to compute various vector particle decay quantities in order to describe leptonic as well as hadronic decay signatures for experimental searches.

\end{abstract}

\maketitle
\flushbottom
\newpage
\tableofcontents

\section{Introduction}
\label{sec:intro}

In recent years MeV-to-GeV scale neutral vector mediators have received a lot of attention being the focus of searches in several  present and future experimental programs. In part this is because they can be involved in the solution of some unsolved conundrums we face today. They have been evoked in association with dark matter models~\cite{Pospelov:2007mp,Kamada:2018zxi,Plehn:2019jeo,Borah:2021jzu,Singirala:2021gok,Holst:2021lzm,Batell:2021snh}, with the muon anomalous magnetic dipole  moment~\cite{Baek:2001kca,Pospelov:2008zw}, with the MiniBooNE excess of electron like events~\cite{Bertuzzo:2018itn,Ballett:2018ynz} and 
to alleviate the reported tension in the Hubble constant~\cite{Escudero:2019gzq}.

Theoretically, these vector bosons appear in connection to extensions of the Standard Model (SM) where the SM gauge group is supplemented by an Abelian  $U(1)_Q$ symmetry. The new gauge coupling $g_Q$, charges ($Q$) and the mass of the vector boson $Z_{Q}$ depend on the particular model realization. 

The vector mediator can be secluded, when only kinetic mixing with the photon is allowed, or can enjoy direct gauge couplings to SM fermions. In the former case, generally dubbed {\it dark photon}, the mediator  
couples universally to all SM charged fermions and ignores neutrinos. In the latter case, the gauge boson may not only interact with all SM fermions but additional particles, vector-like under the SM symmetry group but chiral-like under $U(1)_Q$. Besides, a judicious choice for the charges is generally required to enforce anomaly cancellation. 

One can find many limits on the masses and couplings of these particles in the literature. In Refs.~\cite{Ilten:2018crw,Bauer:2018onh} limits on a few $U(1)_Q$ models for a wide range of masses (from 2 MeV to 90 GeV) were derived or recasted from experimental searches for {\it dark photons}. Most of these searches rely on the mediator decays into leptons, either electrons, muons or neutrinos. In some models, the branching ratios into leptons indeed dominate. Nevertheless, for baryophilic vector bosons decays into hadronic final states might have a large share of the total decay width.   
Especially for a $Z_{Q}$ with a mass in the MeV-to-GeV range, these limits fall in the domain of nonperturbative QCD. Hence, it is important to make sure the hadronic resonances that play an important role in determining the experimental bounds in this region are well described. The main purpose of this paper is to improve this description and provide, for the first time, an almost complete set of $Z_Q$ decays into arbitrary leptonic and hadronic final states. This is of consequence as one can, misguided by an incomplete or incorrect theoretical description of the data, exclude regions that are still allowed and perhaps hinder the imminent discovery of a new weak force. Besides, present bounds and future predictions for vector mediator models could, in principle, be complemented by hadronic signature searches.

In order to obtain reliable predictions in this low mass region we use a data driven approach fitting $e^+e^-$ cross-section data using the meson dominance (VMD) model of chiral perturbation theory. Under this model assumptions we can calculate the decay widths and branching ratios of the new $Z_{Q}$ mediator into hadrons by considering  its direct mixing to the dominant vector mesons $\rho$, $\omega$ and $\phi$. A similar approach  was also used in~\cite{Ilten:2018crw,Bauer:2018onh}, but here we improve their implementation in several ways.

We explicitly calculate the $Z_{Q}$ width to specific hadronic final states following the same procedure outlined in \cite{Plehn:2019jeo} fitting the available $e^+ e^-$ data using \verb!IMinuit! \cite{hans_dembinski_2021_5561211}. Many of those fits are based on state-of-the-art hadronic current parametrizations of $e^+e^-$ annihilation processes~\cite{Rodrigo:2001kf,Czyz:2017veo}, and all fits are updated to the most recent data. Once the hadronic currents for numerous mesonic final states are parametrized and the fit values are fixed, we can couple the weak force to all individual currents.  We include several new hadronic channels with respect to~\cite{Ilten:2018crw}, especially in the region  where there are excited states of the $\rho,\omega,$ and $\phi$ above 1 GeV. The results for the hadronic decays of the new $Z_Q$ mediator are provided in the python package \textsc{DeLiVeR} that is available for public use on GitHub at \href{https://github.com/preimitz/DeLiVeR}{https://github.com/preimitz/DeLiVeR} with a jupyter notebook tutorial.

This paper is organized as follows. In section~\ref{sec:theo} we introduce a class of baryophilic models that we use throughout the paper. The particular couplings to quarks determines the $Z_Q$ decays into light hadrons as described in section~\ref{sec:Zdecays}. In section~\ref{sec:HadCalc} we discuss how the description of those decays can be improved compared to previous calculations by using the VMD approach with only very little theoretical assumptions. The impact this different approach has on the hadronic widths, the branching ratios, and on the reach of  present and future experimental searches for $Z_Q$ vector particles is part of section~\ref{sec:results}. Our conclusion and outlook is presented in section~\ref{sec:conc}.

\section{General Theoretical Framework}
\label{sec:theo}

We will consider extensions of the SM where a new vector boson $Z_Q$ acquires a mass $m_{Z_Q}$ after the spontaneous symmetry breaking of an extra gauged $U(1)_Q$ symmetry~\footnote{We will not specify the scalar sector of the model as it is not needed for our purposes. Note, however, that if the scalar that breaks $U(1)_Q$ is also charged under the SM symmetry group, mass mixing will also be  present. See appendix~\ref{appx:framework} for more details.}.
As it is well known, even if not present at tree-level, kinetic mixing between two $U(1)$ field strength tensors can be generated at loop-level if there are particles charged under both gauge groups~\cite{Holdom:1985ag}. So we will consider the following 
renormalizable Lagrangian allowed by the $SU(2)_L\times U(1)_Y \times U(1)_Q$ gauge symmetry

\be
\mathcal{L}_{\rm gauge} \supset - \frac{1}{4} \hat F_{\mu \nu} \hat F^{\mu \nu}  - \frac{1}{4} \hat Z_{Q \mu \nu} \hat Z_Q^{\mu \nu} - \frac{\epsilon}{2 \cos \theta_W} \hat Z_{Q \mu \nu} \hat F^{\mu \nu} ,
\label{eq:Lgauge}
\ee
with a kinetic mixing of the hypercharge and the $Q$-charge field strength tensors, $\hat{F}_{\mu \nu}=\partial_\mu \hat{F}_\nu - \partial_\mu \hat{F}_\mu$ 
and $\hat{Z}_{Q\mu \nu}=\partial_\mu \hat{Z}_{Q\nu} - \partial_\mu \hat{Z}_{Q\mu}$, respectively. We parameterize 
this mixing by $\epsilon/(2 \cos \theta_W)$ for convenience. 

Considering $\epsilon \ll 1$, we can rotate $\hat{F}$ and $\hat Z$ as
(see appendix~\ref{appx:framework} for details)
$$\hat F_\mu \to F_\mu - \frac{\epsilon}{\cos \theta_W} Z_{Q\mu} \quad \quad {\rm and} \quad \quad \hat Z_{Q\mu} \to Z_{Q\mu}\, ,$$
in  order to define gauge bosons with canonical kinetic terms.
This rotation will also impact the neutral bosons interaction 
Lagrangian so that the relevant terms involving the new physical $Z_Q$ boson are

\be
\mathcal{L}_{\rm int}^{0} \supset  e \epsilon J^\mu_{\rm em} Z_{Q\mu} - g_Q J_Q^\mu Z_{Q\mu} \, ,
\label{eq:Lint}
\ee
where $e=g \sin \theta_W$ is the electric charge, $g$ and $g_Q$ are, respectively, the $SU(2)_L$ and $U(1)_Q$ coupling constants and $\theta_W$ is the SM weak mixing angle. As usual 

\be
 J^\mu_{\rm em} = \sum_{f} \bar{f} \gamma^\mu \, q^f_{\rm em} \, f\, ,
 \ee
is the SM electromagnetic current, $q^f_{\rm em}$ is the fermion $f$ 
electric charge in units of $e$, and 
\be
 J^\mu_Q = \sum_{f} \bar{f} \gamma^\mu \, q^f_{Q} \, f\, ,
 \ee
is the new vector current, with $q^f_{Q}$ being the $Q$-charge of fermion $f$. If only the first term is present in eq.~\eqref{eq:Lint}, i.e. if $q^f_{Q}=0$ for all fermions, the boson will couple universally to all charged fermions and we will refer to it  as the {\it dark photon} $Z_{\gamma}$. We will assume $e \epsilon \ll g_Q$, so when charges are present we will neglect the kinetic mixing contribution.

Because our main focus here are the hadronic modes for a light $Z_{Q}$ with $m_{Z_Q}$ in the MeV-to-GeV range, we will consider 
a class of anomaly-free baryophylic models where only three right-handed neutrinos were introduced to the particle content of the SM~\cite{Araki:2012ip}. 
The symmetry generator for these models can be written as
\be
Q = B - x_e L_e - x_\mu L_\mu -(3-x_e-x_\mu)L_\tau \,, 
\label{eq:charges}
\ee
where $B$ is the baryon number and $L_e, L_\mu$  and $L_\tau$ are lepton family number operators. To compare with previous works, we will also present our results for the $B$ model.
In table~\ref{tab:models}, we list the 
models we will use in this work. 

\begin{table}[htb]
\def\arraystretch{2.0}
    \centering
    \begin{tabular}{|c|c|c|c c c c| }
    \hline
    \hline
        $x_e$ & $x_\mu$ & $Q$ & \multicolumn{4}{c|}{ $q^f_Q$}  \\
         & &  & quarks & $e/\nu_e$ & $\mu/\nu_\mu$ & $\tau/\nu_\tau$\\
        \hline \hline
         1 & 1 & $B-L$ & $\frac{1}{3}$ & -1 & -1 & -1\\
         \hline
         3 & 0 & $B-3L_e$ & $\frac{1}{3}$ & -3 & 0 & 0\\
         \hline
         0 & 3 & $B-3L_\mu$ & $\frac{1}{3}$ & 0 & -3 & 0\\
         \hline
         0&0 & $B-3L_\tau$ & $\frac{1}{3}$ & 0 & 0 & -3\\
         \hline
         1 & 0 & $B-L_e-2L_\tau$ & $\frac{1}{3}$ & -1 & 0 & -2\\
         \hline
         0 & 1 & $B-L_\mu - 2 L_\tau$ & $\frac{1}{3}$ & 0 & -1 & -2\\
         \hline
         -- & -- & $B$ & $\frac{1}{3}$ & 0 & 0 & 0\\
         \hline \hline
    \end{tabular}
    \caption{Symmetry generators and fermion charges for the models considered in this work.}
    \label{tab:models}
\end{table}

\section{On the Decays of $Z_Q$}
\label{sec:Zdecays}

In the mass range of interest of this paper, a $Z_{Q}$ can decay into charged or neutral leptons as well as into light hadrons, if  kinematically allowed. In the following, we describe its partial decay widths into these channels.

\vglue 0.1 cm 
\paragraph{Leptonic Decays:}
The $Z_{Q}$ partial decay width into a pair of leptons  is
given by
\be \label{eq:Gammaffbar}
\Gamma (Z_Q \to \ell \bar \ell) = \frac{ C_{\ell}(g_Q q_Q^\ell)^2}{12 \pi} m_{Z_Q} \left( 1 + 2\,  \frac{m_\ell^2}{m_{Z_Q}^2}\right) \sqrt{1- 4\frac{m_\ell^2}{m_{Z_Q}^2} }\, ,
\ee
where $m_\ell$ is the lepton mass, $C_{\ell}=1 \, (1/2)$ for $\ell = e, \mu$ ($\nu_e, \nu_\mu, \nu_\tau$), $g_Q$ is the $U(1)_Q$ gauge coupling 
and $q_Q^\ell$ the corresponding lepton charge of the model according to table~\ref{tab:models}. For the {\it dark photon} we have to replace $g_Q$ with $e \epsilon$ and $q_Q^{\ell}$ with $q_{\rm em}^{\ell}=-1$ for all charged leptons as neutrinos do not couple to $Z_\gamma$.

When $m_{Z_Q}< 2 \, m_e$, the new boson $Z_Q$ can also decay into three photons. The decay width for this process can be found in \cite{Seo:2020dtx}. Nevertheless, the partial decay width for $Z_Q\to 3\gamma$ is negligibly small for the models and mass range of interest in this paper and hence, we refrain from including it into our calculations.
\vglue 0.2 cm
\paragraph{Hadronic Decays:} The region $0.5 \lesssim m_{Z_Q}/{\rm GeV} \lesssim 2$ is plagued by hadronic resonances and perturbative 
QCD does not provide a reliable way to evaluate  vector boson decays  
into hadrons, so here one has to, instead, turn to chiral perturbation theory~\cite{Scherer:2002tk}. We will use the so-called  
vector meson dominance model~\cite{Sakurai:1960ju,Kroll:1967it,Lee:1967iv}, which  successfully describes 
$e^+e^-$ annihilations into hadrons and has been also 
applied more recently to BSM physics~\cite{Tulin:2014tya,Ilten:2018crw}, to estimate $Z_{Q} \to \mathcal{H}$, where $\mathcal{H}$ is a hadronic final state made of light quarks, via mixing with QCD vector mesons.

Let's explain briefly how VMD works to describe low-energy QCD.
In a nutshell, in the context of SM interactions, VMD  splits the electromagnetic light quarks current into three components, 
the isospin $I=0,I=1$ and the strange quark currents, and identifies them, respectively, with the vector mesons $\omega, \rho$ and $\phi$~\cite{OConnell:1995nse}. The same result can be obtained by incorporating dynamical gauge fields $V_\mu$ 
of a local hidden symmetry $U(3)_V$~\cite{Bando:1984ej,Fujiwara:1984mp,Kramer:1984cy,Bando:1985rf,Bando:1987br} into the chiral Lagrangian~\cite{Fujiwara:1984mp,Scherer:2002tk}. Linear combinations of these gauge fields will then describe the vector mesons. The vector mesons subsequently interact with other vector mesons $V'$ and pseudoscalar mesons $P$ through the anomalous Wess-Zumino-Witten (WZW) $VV'P$ interactions~\cite{Fujiwara:1984mp,Kramer:1984cy,Bando:1987br}.

With the most prominent example being the SM photon, $U(1)$ gauge symmetric fields, such as $Z_{Q\mu}$, enter this pure QCD Lagrangian as external fields through the covariant derivative of the pseudoscalar Goldstone matrix of the chiral Lagrangian~\cite{Scherer:2002tk}. Additional WZW terms~\cite{Wess:1971yu,Witten:1983tw} are constructed to fully describe the meson sector such as, for example, the $\pi^0\to\gamma\gamma$ decay. Whereas in the low-energy limit those U(1) gauge fields interact directly with the pseudoscalar mesons, they dominantly mix with vector mesons in the hadronic resonance region. Hence,  we only have to specify the vector meson-gauge field mixing term. Its most general form is given by\footnote{For alternative definitions see~\cite{OConnell:1995nse}.}
\begin{align}
\mathcal{L}_{VZ_Q}=2\, g_{Q}  Z^\mu_Q{\rm Tr}\left[V_\mu Q^{f}\right]\, ,
\label{eq:mixing}
\end{align}
with $V^\mu=T^a V^{a,\mu}$, where $T^a$ are $U(3)$ generators.
In our case $Q^{f}$ is a diagonal matrix with entries equal to the $U(1)_Q$ charges $q^{u,d,s}_Q$.  For  the {\it dark photon} $Z_{\gamma}$ one can simply take  $g_Q \to e \epsilon$ and $q^f_{Q} \to q^f_{\rm em}$.

 The observed vector mesons of the SM are given by
\begin{align}
   &\rho : \; \rho^{\mu} T_\rho =\rho^{\mu} \frac12 {\rm diag}(1,-1,0)\, ,\notag\\
   &\omega : \;  \omega^{\mu} T_\omega =\omega^{\mu}\frac12 {\rm diag}(1,1,0)\, ,\notag\\
   &\phi : \;  \phi^{\mu}T_\phi =\phi^{\mu}\frac{1}{\sqrt{2}}{\rm diag}(0,0,1)~.
   \label{eq:mesons}
\end{align}
Once the vector mediator $Z_{Q}$ has converted into a SM vector meson, the $VV'P$ interactions, e.g.~the $\rho\omega\pi$ vertex, determine their decays. These interactions are encoded in QCD form-factors $F(q^2)$. The low-energy limit of chiral perturbation theory is always recovered in the VMD model by making $F(q^2)\to 1$ for $q^2\to 0$.

\begin{figure}[h]
\begin{center}
\includegraphics[width=1.0\textwidth]{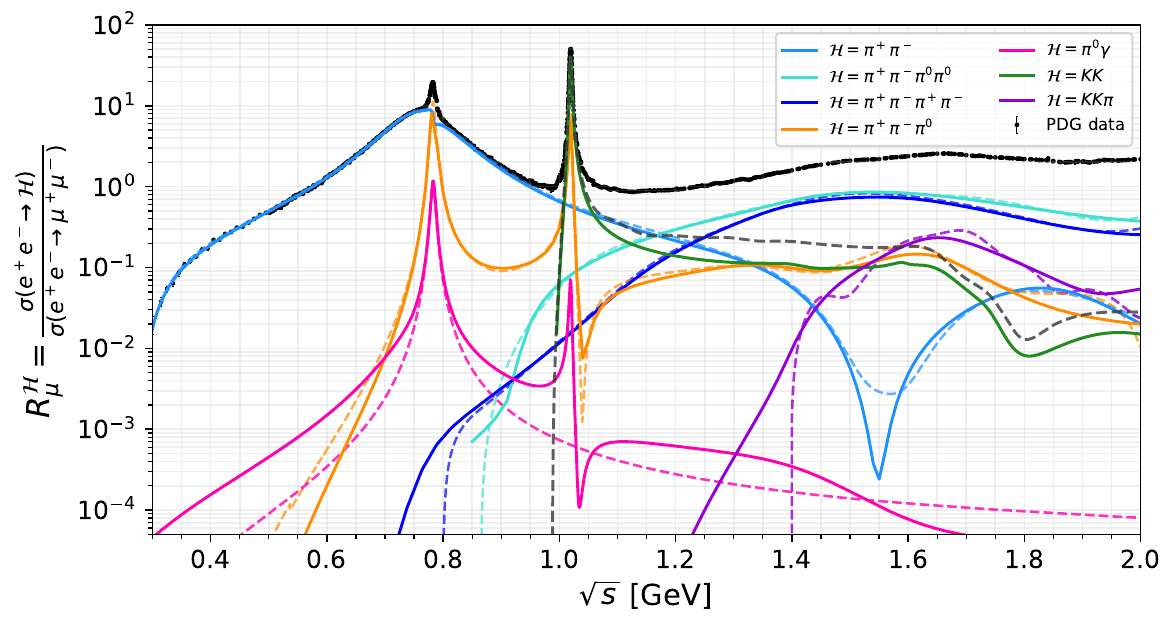}
\end{center}
\vglue -0.8 cm
\caption{\label{fig:xsec} Cross-sections for the dominant $e^+e^-\to \mathcal{H}$ channels, normalized by the $e^+e^-\to \mu^+\mu^-$ cross-section. The solid (dashed) lines indicate results obtained 
in this work (taken from \textsc{DarkCast}~\cite{Ilten:2018crw}). 
The data (black points) was taken from the Particle Data Group compilation (PDG)~\cite{ParticleDataGroup:2020ssz}.
See text for discussions on the differences.} 
\end{figure}

All form-factors can be obtained from fits to $e^+e^- \to \mathcal{H}$ data. The cross-section results are typically displayed as the ratio  over the muonic annihilation channel as
\begin{align}
    R_\mu^{\mathcal{H}} \equiv \frac{\sigma(e^+e^-\to \mathcal{H} )}{\sigma(e^+e^-\to\mu^+\mu^-)}\, .
\label{eq:RSM}
\end{align}
This common rescale of the results for vector portal models is 
justified since initial state dependencies  cancel in the above ratio. 
As in the {\it dark photon} model, the coupling structure is inherited from the SM photon with a proportionality factor $\epsilon$, we can hence model the {\it dark photon} decay widths simply by directly rescaling the experimentally known   ratios $R_\mu^\mathcal{H}[\text{\small exp}] \equiv \sigma(e^+ e^- \to \mathcal{H})/ \sigma(e^+e^- \to \mu^+ \mu^-)\vert_\text{exp}$ as 
\begin{align}
\Gamma_{Z_{\gamma}\to \mathcal{H}}=\Gamma_{Z_\gamma \to \mu^+\mu^-} R_\mu^\mathcal{H}[\text{\small exp}].
\label{eq:widthDP}
\end{align}

Although this strategy works well for the {\it dark photon}, it cannot be employed anymore when dealing with vector mediators with a coupling structure that is not proportional to the SM photon-quark one. In this scenario the couplings to SM vector mesons need to be determined  by eq.~\eqref{eq:mixing}. For instance, in all the models of interest in this paper,  $Z_Q$ couples to $B$, so the quark $U(1)_Q$ charge matrix takes the form $Q^f={\rm diag}(1/3,1/3,1/3)$. In this case the trace for the $\rho$ meson will be zero and hence, only the $\omega$ and $\phi$ mesons will contribute to describe the $Z_Q$ decay into hadrons in $R_\mu^\mathcal{H}$.

Therefore, for generic $U(1)_Q$ models, an accurate division of the hadronic channels into their $\rho$, $\omega$ and $\phi$ contributions is of extreme importance in order to obtain the correct description of the hadronic decay widths. In previous studies the VMD approach has been employed with many simplifications and considering a limited number of hadronic channels~\cite{Ilten:2018crw}. These approximations propagate to the width and branching ratio calculations, and can even affect the final experimental bounds in the model parameter space. Next we present a more complete 
and robust evaluation of various hadronic contributions.

\section{Improvements in the Hadronic Calculation}
\label{sec:HadCalc}

Here we describe the improvements we have implemented in the 
calculation of the widths and branching ratios of $Z_{\gamma,Q}$ into light hadrons and compare our results with what was used by ref.~\cite{Ilten:2018crw} and is included in the \textsc{DarkCast} code. 

\vglue 0.2 cm

\paragraph{Calculation of $\sigma(e^+ e^- \to \mathcal{H})$:} 
Instead of using the ratio of the total hadronic over muonic annihilations $R_\mu^\mathcal{H}$ in $e^+e^-$-processes to estimate the hadronic widths of $Z_{\gamma,Q}$, as in the above mentioned previous work, we have explicitly calculated the  
individual cross-sections $\sigma(e^+ e^- \to \mathcal{H})$ which enter eq.~(\ref{eq:RSM}), and contribute to the total hadronic cross-section for the energy range from the pion threshold up to slightly below 2 GeV, using  the VMD effective method and experimental data to fit the parameters of the model.

In order to precisely determine the $\rho$-like, $\omega$-like and $\phi$-like contributions to a particular hadronic channel, we parametrize each individual channel playing a part in $e^+e^-\to {\rm hadrons}$ in terms of its underlying vector meson dominance. 

The matrix-element for a given process  $e^+e^-\to \mathcal{H}$ can be written as
\begin{align}
    \mathcal{M}_{e^+e^-\to \, \mathcal{H}}=\mathscr{L}_\mu J^\mu_\mathcal{H}\, ,
    \label{eq:MEee}
\end{align}
where $$\mathscr{L}_\mu=e^2\frac{g_{\mu\nu}}{s}\bar{v}(k_{e^+})\gamma^\nu u(k_{e^-})\, ,$$ is the leptonic current, $J^\mu_\mathcal{H}$ is the hadronic current and $\mathcal{H}$ is one of the individual final state configurations $\mathcal{H}=2\pi,3\pi,K\bar{K},...$ we consider here. 
The hadronic current, which includes a form-factor $F_\mathcal{H}$, 
depending on $\mathcal{H}$, can be written as
\begin{align}
    J^\mu_{P_1 P_2}&=-(p_1-p_2)^\mu F_{P_1 P_2}(q^2), &\quad J^\mu_{P\gamma}&=\varepsilon^{\mu\nu\rho\sigma}q_\nu \varepsilon_{\gamma,\rho}p_{\gamma,\sigma} F_{P\gamma}(q^2),\\
    J^\mu_{VP}&=\varepsilon^{\mu\nu\rho\sigma}q_\nu\varepsilon_{V,\rho}p_{P,\sigma}F_{VP}(q^2), &\quad J^\mu_{P_1 P_2 P_3}&=\varepsilon^{\mu\nu\rho\sigma}p_{1,\nu}p_{2,\rho}p_{3,\sigma}F_{P_1 P_2 P_3}(p_1,p_2,p_3)\, ,
\end{align}
where $P_{(1,2,3)}, V, \gamma$ indicate, respectively, the presence of a pseudoscalar meson, a vector meson, or a photon in the final state. 
The corresponding momenta are labeled accordingly. The photon and vector meson polarizations are given by $\varepsilon_{\gamma/V,\mu}$ and $\varepsilon^{\mu\nu\rho\sigma}$ is the antisymmetric Levi-Civita tensor. For the pseudoscalar-pseudoscalar current, the pseudoscalar-photon current and the pseudoscalar-vector current, we have $q=p_1+p_2$, $q=p_{P}+p_{\gamma}$ and $q=p_{P}+p_{V}$, respectively.

For channels with two pseudoscalars and one vector meson, as in the case of $\omega\pi\pi$ and $\phi \pi \pi$, we refrain from parametrizing the hadronic current in terms of intermediate substructures like $\omega f_0\to\omega\pi\pi$ due to dissenting data observations~\cite{Aubert:2007ef,Lees:2018dnv}. Hence, we assume a point-like interaction and write the hadronic current as
\begin{align}
    J^\mu_{VP_1 P_2}=\left(g^{\mu\nu}-\frac{q^\mu q^\nu}{q^2}\right)\varepsilon^*_{V,\nu}F_{VP_1 P_2}(q^2)\, ,
    \label{eq:VPP}
\end{align}
with $q=p_V+p_1+p_2$.
For channels with more than 3 final states, we directly take expressions from the literature as given in table~\ref{tab:new_channels}.

In order to calculate the decay width of a vector mediator, we simply replace the leptonic current by the polarization vector of the mediator $L_\mu \to \varepsilon_{\mu}(Z_Q)$ 
to obtain the matrix element for the decay, so 
\be
\mathcal{M}_{Z_Q\to \mathcal{H}}=\varepsilon_{\mu}(Z_Q) \sum_V r(V) J^\mu_{\mathcal{H}}(V)\, ,
 \quad \quad r(V)=\frac{ g_{Q}{\rm Tr}\left[T_V Q^{f}\right]}{{\rm Tr}\left[T_V Q^{\rm em}\right]} \, ,
\ee
with the factor $r(V)$ rescaling the photon-meson coupling to the mediator-meson coupling with the vector meson resonance $V$, in this case $V=\rho,\omega,\phi$ with generators $T_V$ as given in eq.~(\ref{eq:mesons}).

The dependence on the vector meson resonances $\rho, \omega$ and $\phi$ will appear in the form-factors $F_\mathcal{H}$. The dominant vector mesons for a particular channel can be identified using isospin-symmetry assumptions and G-parity conservation. The particular form of these form-factors can be found in \cite{Plehn:2019jeo} and in appendix~\ref{appx:newfit}.

In this work, we include the cross-sections for the four most important hadronic contributions close to the $\rho,\omega,$ and $\phi$ masses as well as the $4\pi$ and $KK\pi$ channels that are already part of \textsc{DarkCast} 
(see table~\ref{tab:old_channels}), but also consider several new hadronic channels (see table~\ref{tab:new_channels}) using recent data for the parametrizations. Some of those additional new channels are taken from \cite{Plehn:2019jeo}, and are complemented by new fits to other channels not considered before in the energy range closer to $\sim 2$ GeV. Table \ref{tab:new_channels} summarizes all additional channels and specifies the vector resonances used in the fit as well as possible final state configurations.

\vspace{5pt}
\paragraph{Improvements on the Description of the Dominant Low-Energy Hadronic Modes:} 

For energy ranges around the ground state vector meson masses, the final states $\pi^0\gamma$, $\pi^+\pi^-$, $\pi^+\pi^-\pi^0$, $KK$ and $KK\pi$ dominate the cross-section. For higher energies we also include the contribution from $e^+ e^- \to 4\pi$. 
Those channels are very precisely measured and have been also considered by \textsc{DarkCast}. In table~\ref{tab:old_channels}, we list the assumptions for resonant contributions and its differences to \textsc{DarkCast}, the data used, and references for the parametrizations and fits.

In figure~\ref{fig:xsec}, we show our results (solid lines) for these modes and compare them to the state-of-the-art results from \textsc{DarkCast} (dashed lines). Whereas the results are similar around the $\rho$ and $\omega$ masses, channels including the $\phi$ meson give different results. Below, we summarize the main improvements and explain the differences for these channels introduced in our work:

\begin{itemize}
    \item In the $\pi^0\gamma$ channel, besides the $\omega$-like components 
    we include a $\phi$ and a small  $\rho$ contribution. The $\phi$, in particular, accounts  for a second peak around its mass
    near 1~GeV (see pink solid line in figure~\ref{fig:xsec}) and for the 
broadening  of the $\omega$ peak. Especially in the low-energy limit, below $\simeq 0.6$~GeV, this might have some significant effect if no other hadronic states contribute to the overall decay width of the vector mediator. In the particular case of a $B$ vector boson model, this modification will visibly affect branching ratios, and hence, may modify  model limits.
    
    \item Regarding the $KK$ channel, we fit both the charged $K^+K^-$ and neutral $\bar{K}^0K^0$ components separately, instead of taking $KK=2\; K^+K^-$ as in the  \textsc{DarkCast} code. The latter calculation leads to the overestimation of the total $KK$ cross-section (see dashed green line in figure~\ref{fig:xsec}). We also consider the contributions from $\rho$-like, $\omega$-like and $\phi$-like mesons, and not only from $\phi$. The inclusion of these other mesons may have an important impact for models that do not couple to the $\rho$ current, such as the baryophilic $Z_Q$ models considered here.
    
    \item Finally, the $KK\pi$ channel can be decomposed into three components  
    $\mathcal{H}= K^0 K^0\pi^0$, $K^+K^-\pi^0$ and $K^\pm K^0\pi^\mp$. In \textsc{DarkCast} these components are not considered individually. Instead, only the isoscalar component of the $KK\pi$ channel has been taken into account. The isoscalar and isovector contributions can be extracted from the sub-process $e^+ e^- \to{K^*(892)K}$, e.g.~in the analysis of $e^+e^-\to K^\pm K^0\pi^\mp$~\cite{BaBar:2007ceh}. However, this is a two-body process, and therefore has different kinematics compared to a three-body final state. So in order to correctly describe the kinematics of $KK\pi$ we need to make the decomposition into the three final states. Moreover, we take into account the $\rho$-like and  $\phi$-like contributions, while  \textsc{DarkCast} assigns the whole $[KK \pi]_{I=0}$ as a $\phi$-like channel.
    The difference between these calculations can be seen in figure~\ref{fig:xsec} (purple lines).
    
\end{itemize}

\begin{table}[t!]
\centering
\begin{tabular}{l|ccccc}
\hline
channel & resonances & data & parametrization & fit & possible final states\\ \hline
$\pi\gamma$ & $\bm{\rho},\omega, \bm{\omega'}, \bm{\omega''}, \bm{\phi}$ & \cite{SND:2016drm} & \cite{SND:2016drm} & \cite{SND:2016drm} & $\pi\gamma$ \\
$\pi\pi$ & $\rho, \rho', ...$ & \cite{KLOE:2008fmq,BaBar:2009wpw,BaBar:2012bdw} &  \cite{Czyz:2010hj} & \cite{Czyz:2010hj} & $\pi\pi$ \\
$3\pi$ & $\bm{\rho}, \bm{\rho}'', \omega, \omega',\omega'',\phi$ & \cite{BaBar:2004ytv} & \cite{Czyz:2005as} & \cite{Czyz:2005as} & $3\pi$\\
$4\pi$ & $\rho,\rho',\rho'',\rho'''$ & \cite{BaBar:2012sxt,BaBar:2017zmc} & \cite{Czyz:2008kw} & \cite{Plehn:2019jeo} & $4\pi$\\
$KK$ & $\bm{\rho},...,\bm{\omega},...,\phi,...$ & \cite{CLEO:2005tiu,Achasov:2000am,Achasov:2006bv,Mane:1980ep,CMD-3:2016nhy,BaBar:2014uwz,CMD-2:2008fsu,BaBar:2013jqz,BaBar:2015lgl,Achasov:2016lbc} & \cite{Czyz:2010hj} & \cite{Plehn:2019jeo} & $KK$ \\
$KK\pi$ & $\bm{\rho},\bm{\rho'},\bm{\rho''},\phi,\phi',\phi''$ & \cite{BaBar:2007ceh,BaBar:2017nrz,Achasov:2017vaq,Bisello:1991kd,Mane:1982si} & \cite{Plehn:2019jeo} & \cite{Plehn:2019jeo} & $KK\pi$ \\
\hline
\end{tabular}
\caption{Dominant hadronic processes included in this work as well as in the \textsc{DarkCast} code~\cite{Ilten:2018crw}. We specify the resonances included in the first but not in the latter in boldface~\footnote{\textsc{DarkCast} takes into account higher resonances in an 
approximate way by adding a non-resonant background function to mimic the shape of the data, whereas we stick to the VMD assumption and calculate each channel by considering resonance contributions.} and denote channels where a tower of vector meson resonances was considered with `...'. As possible final states we consider low-energy pseudoscalar mesons, $\pi$ and $K$, as well as photons.}
\label{tab:old_channels}
\end{table}

\begin{table}[t!]
\centering
\begin{tabular}{l|ccccc}
\hline
channel & resonances & data & parametrization & fit & possible final states\\ \hline
$\eta\gamma$ & $\rho,\rho',\omega,\phi$ & \cite{Achasov:2006dv} & \cite{Achasov:2006dv} & \cite{Achasov:2006dv} &$3\gamma$, $3\pi\gamma$,...\\
$\eta\pi\pi$ &  $\rho,\rho',\rho''$ & \cite{Achasov:2017kqm,TheBABAR:2018vvb} & \cite{Czyz:2013xga} & \cite{Plehn:2019jeo} & $2\pi2\gamma,5\pi$,...\\
$\omega\pi\to\pi\pi\gamma$ & $\rho,\rho',\rho''$ & \cite{Achasov:2016zvn} & \cite{Achasov:2016zvn} & \cite{Achasov:2016zvn} & $2\pi\gamma$\\
$\omega\pi\pi$ & $\omega''$& \cite{Akhmetshin:2000wv,Aubert:2007ef,Lees:2018dnv} & new & new &$5\pi, 3\pi\gamma$\\
$\phi\pi$ & $\rho,\rho'$ & \cite{BaBar:2007ceh,TheBABAR:2017vgl} & \cite{Plehn:2019jeo} & \cite{Plehn:2019jeo} &$2K\pi$, $4\pi$,...\\
$\eta' \pi\pi$ & $\rho'''$ & \cite{Aubert:2007ef} & \cite{Czyz:2013xga} & \cite{Plehn:2019jeo} &  $4\pi2\gamma$,...\\
$\eta\omega$& $\omega',\omega''$ & \cite{Achasov:2016qvd} & \cite{Plehn:2019jeo} & \cite{Plehn:2019jeo} & $2\pi2\gamma,6\pi,...$\\
$\eta\phi$  & $\phi',\phi''$ & \cite{BaBar:2007ceh,Achasov:2018ygm} & \cite{Plehn:2019jeo}  & \cite{Plehn:2019jeo} &$KK2\gamma, KK3\pi,...$\\
$p\bar{p}/n\bar{n}$ & $\rho,\rho',...,\omega,\omega'$,... & \cite{Lees:2013ebn,Ablikim:2015vga,Pedlar:2005sj,Delcourt:1979ed,Castellano:1973wh,Antonelli:1998fv,Armstrong:1992wq,Ambrogiani:1999bh,Andreotti:2003bt,Punjabi:2005wq,Puckett:2011xg,Gayou:2001qt,Puckett:2010ac,Puckett:2017flj,Pospischil:2001pp,Plaster:2005cx,Geis:2008aa,Andivahis:1994rq} & \cite{Czyz:2014sha} & \cite{Plehn:2019jeo} &$p\bar{p}/n\bar{n}$\\
$\phi\pi\pi $ & $\phi',\phi''$ & \cite{BaBar:2011btv,Belle:2008kuo} & new & new & $K K \pi \pi$ \\
$K^*(892) K \pi $ & $\rho'',\phi'$ & \cite{BaBar:2011btv,BaBar:2017pkz} & new & new & $K K \pi \pi$ \\
$6 \pi $ & $\rho'''$ & \cite{BaBar:2006vzy} & \cite{BaBar:2006vzy} & new & $6 \pi$ \\
\hline
\end{tabular}
\caption{Additional processes included in this work that are not present in the \textsc{DarkCast} code~\cite{Ilten:2018crw}. We denoted channels where a tower of vector meson resonances was considered with `...'.  For the cases where the parametrization and fit are marked as `new', we provide details in appendix~\ref{appx:newfit}.}
\label{tab:new_channels}
\end{table}

\vspace{5pt}
\paragraph{Higher resonance effects:} The even more challenging energy region starts above the $\phi$ mass and includes processes involving excited states of the vector mesons 
$\rho', \omega', \phi'$. The only channel that is rather straightforward to be implemented is the $\rho$ meson dominated $4\pi$ channel with form factors as given in ref.~\cite{Czyz:2008kw} (see navy blue and cyan lines in figure~\ref{fig:xsec}). Other processes, especially vector mediator decays to currents involving $\omega$ and $\phi$ contributions, are only poorly described in the literature. We introduce a large amount of new channels in order to accurately describe the region above $\gtrsim 1.5~\GeV$. To reduce the vast amount of possible final states, we identify common substructures of some channels. For example, the channel $\eta\omega$  can produce $2\pi2\gamma$ and $6\pi$ final states\footnote{Even though the $\eta\omega\to 6\pi$ contribution is expected to be subdominant~\cite{BaBar:2006vzy}.}, whereas $\omega\pi\pi$ can  contribute to $5\pi$ and $3\pi\gamma$. 
 
\begin{figure}[h]
\begin{center}
\includegraphics[width=1.0\textwidth]{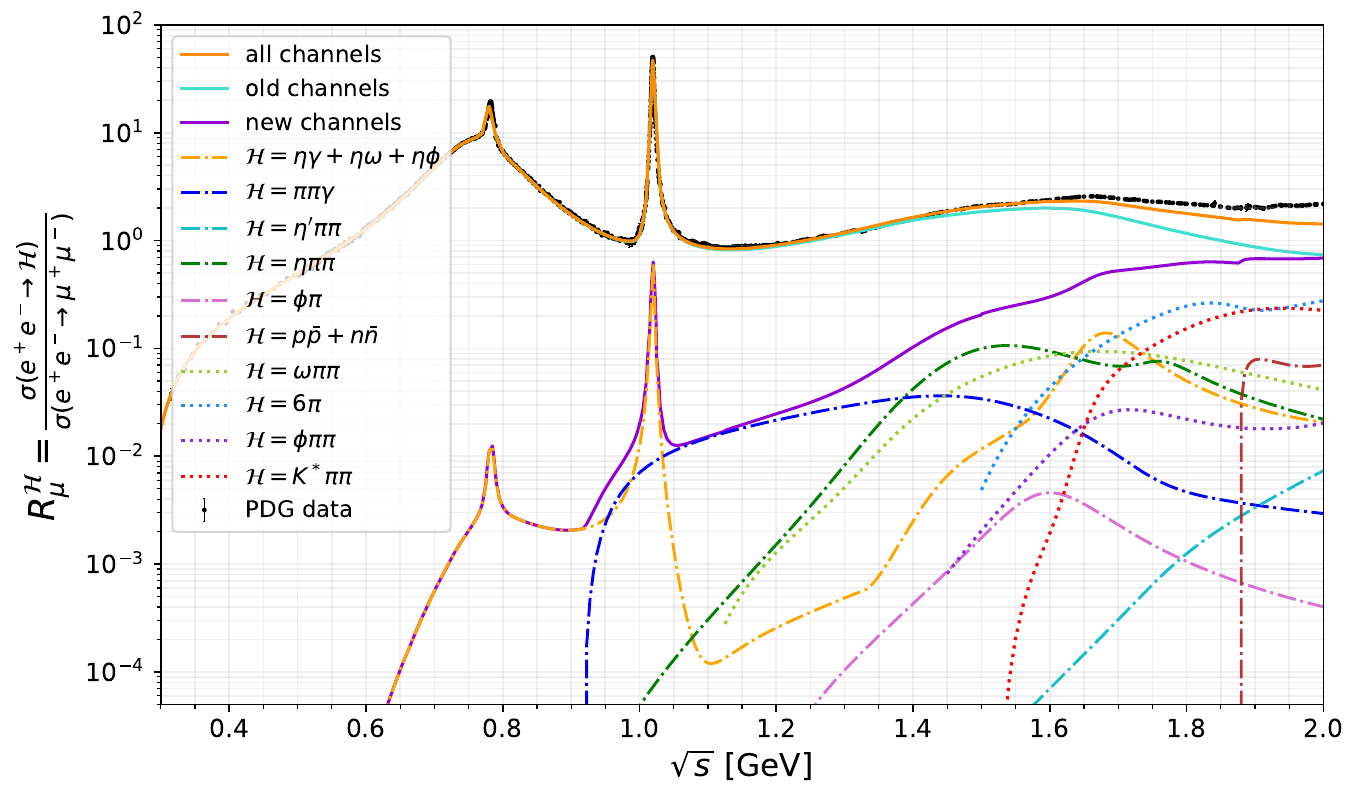}
\end{center}
\vglue -0.8 cm
\caption{\label{fig:newchannels} 
Same as figure~\ref{fig:xsec} but for the new channels included in this work. The dot-dashed lines indicate the hadronic channels already considered in \cite{Plehn:2019jeo} (but not in \cite{Ilten:2018crw}), while the dotted lines indicate channels we have fitted and included here for the first time. 
The solid lines indicate the total $R_\mu^{\mathcal{H}}$ (summed over all hadronic final states) considering: only the channels shown in figure~\ref{fig:xsec} (cyan), only 
the new channels on table~\ref{tab:new_channels} (purple), the sum of all contributions 
we have calculated (orange).} 
\end{figure}

 All considered channels and some of their possible final state configurations are listed in table~\ref{tab:new_channels}. 
 Including additional channels has a significant effect on the total $e^+e^-\to {\rm hadrons}$ cross-section. As seen in figure~\ref{fig:newchannels}, the sum of the new  contributions to the hadronic cross-section increases up to a level where it contributes as much as the so far considered channels at around $\sqrt{s}\lesssim 2~\GeV$ (purple line). For center-of-mass energies $\sqrt{s} \gtrsim 1.4~\GeV$, the $R_\mu^\mathcal{H}$ line continues to follow the PDG-data to higher energies and captures the effects of excited states of the $\rho, \omega,$ and $\phi$ mesons.
 
 One can also see from figure~\ref{fig:newchannels} that the addition of the new channels $\omega \pi \pi$, $6\pi$, $\phi \pi \pi$ and $K^*K\pi$ are important especially in the region near $2$ GeV where they dominate. The $\omega \pi \pi$ ($\phi \pi \pi$)  channel correspond to a neutral and a charged contribution, $\omega \pi^0 \pi^0$ ($\phi \pi^0 \pi^0$) and  $\omega \pi^+ \pi^-$ ($\phi \pi^+ \pi^-$), respectively. The 6$\pi$ channel can also be split into two components, $3(\pi^+ \pi^-)$ and $2(\pi^+ \pi^- \pi^0)$, while the $K^*K\pi$ can be split into four components, $K^{*0}K^{\pm}\pi^{\mp}$, $K^{*\pm}K^{0}_S\pi^{\mp}$,   $K^{*\pm}K^{\mp}\pi^{0}$ decaying into $K^0_S K^\pm \pi^\mp \pi^0$, and $K^{*0}K^{-}\pi^{+}$ decaying into $K^+ K^- \pi^+ \pi^-$. More details about these channels can be found in appendix \ref{appx:newfit}.

\vspace{5pt}

\begin{figure}[h]
\begin{center}
\includegraphics[width=1.0\textwidth]{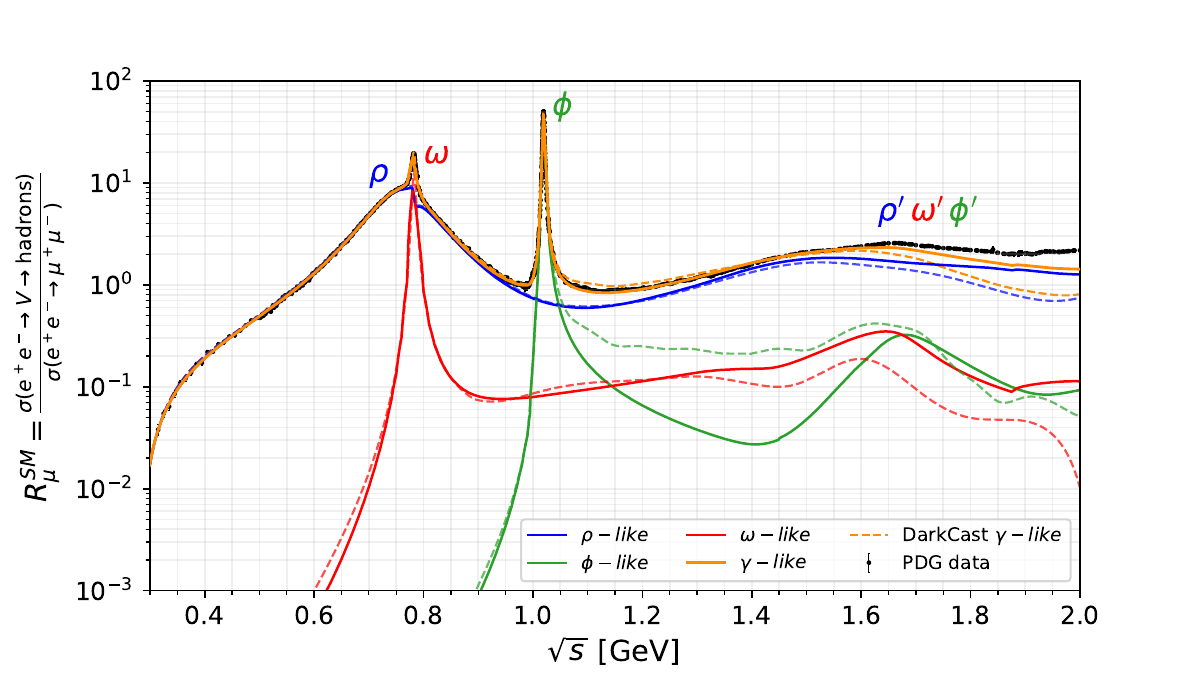}
\end{center}
\vglue -0.8 cm
\caption{\label{fig:contributions} Decomposition of the total hadronic cross-section ratio $R_\mu^{\rm SM}\equiv\sum_\mathcal{H}R_\mu^{\mathcal{H}}$  into $\rho$-, $\omega$- and $\phi$-like contributions for the SM. We also show in orange the total $\gamma$-like contribution.
The dashed lines indicate results obtained with the  \textsc{DarkCast} code~\cite{Ilten:2018crw}. }
\end{figure}

\paragraph{Final $\rho, \omega$, $\phi$ decomposition} In order to calculate decay widths for arbitrary vector mediator models, it is useful to split up the hadronic current in its $\rho, \omega$ and $\phi$ contributions as given in eq.~(\ref{eq:mixing}). The quark coupling matrix $Q^f$ determines if a certain vector meson contribution is present or  absent (${\rm Tr}\left[T_V Q^f\right]=0$). We can clearly see in figure~\ref{fig:contributions} that the different treatment of the $\pi^0\gamma$, $KK$, and $KK\pi$ channels translate into a different $\phi$ contribution above the $\phi$ mass threshold compared to \textsc{DarkCast}. Since we include a lot more channels in the range above $\geq 1.5$~GeV, we also get enhanced $\omega$ and $\rho$ contributions. For vector mediator models with only $\omega$ and $\phi$ couplings, like for example all the $B$-coupled models considered in this paper, this will result in different branching ratios into hadronic final states. 

Due to the fact that in the SM the photon mixes with all vector mesons, in the ideal case we expect the $\gamma$-line to follow the PDG data~\cite{ParticleDataGroup:2020ssz}. As seen in figure~\ref{fig:contributions}, we can accurately describe the $\gamma$-like until around $\sim 1.7$~GeV.  While the $\gamma$-line is almost but not fully overlapping with the $e^+e^-$-data, we have a more solid description of the separate vector meson contributions due to our approach of summing up all dominant meson channels with subsequent $\rho, \omega,$ and $\phi$ vector meson structures.

Especially in the case of the $\omega$ and $\phi$ contributions, the vector meson contributions differ from the calculations of~\cite{Ilten:2018crw} for vector mediator masses above the $\phi$ meson mass, as well as in the low-energy region of the $\omega$ contribution due to differences in the $\pi^0\gamma$ channel. In which way this affects
the branching ratios, limits and predictions will be discussed in section~\ref{sec:results}.
 
\paragraph{Hadron-quark transition} 
 For higher masses than $\gtrsim 1.7$~GeV, we slightly underestimate the $e^+e^-$ total hadronic cross-section due to missing subdominant multi-meson channels.
 Although we have included all the available data of the exclusive channels listed in PDG \cite{ParticleDataGroup:2020ssz}, our results could be improved with better knowledge of the processes and the channels substructures. Also, the inclusion of more data related to final state configurations in the region closer to 2 GeV would improve even more the reach of our $\gamma$-like curve. Possible new channels could be easily added in our approach. Nevertheless, we expect that in that mass range, the annihilation processes slowly transition into perturbative quark production where we have $R_\mu^\mathcal{H} \to R_{\rm em}= N_c\cdot \sum_f (q^f_{\rm em})^2=2$ for the SM with $N_c=3$, $q^u_{\rm em}=2/3$ and $q^{d,s}_{\rm em}=-1/3$.

 In accordance with the PDG~\cite{ParticleDataGroup:2020ssz}, we take 
 \be\label{eq:Rpert}
 R(Q)=R_{\rm em}(1+\delta_{\rm QCD}(Q))\, ,
 \ee
 including QCD corrections $\delta_{\rm QCD}(Q)$ that are described in more detail in the QCD review of~\cite{ParticleDataGroup:2020ssz}.
 As a consequence, due to the lack of sufficient data, the $\gamma$-like curve in figure~\ref{fig:contributions} will be replaced by a perturbative line at $R_\mu^\mathcal{H} \simeq 2$. For the \textit{dark photon} model this transition is made at 1.7 GeV, whereas for $B$-coupled models it is at 1.74 GeV. These specific values for the threshold energies were chosen in the intersection between the perturbative quark width, calculated using eq.~(\ref{eq:Rpert}) and the width to muons, and the hadronic width, such that the transition can be done smoothly.
 
\paragraph{Error estimate:} The uncertainties in our calculation of hadronic decays of light vector mediators emerge from uncertainties from the fits to electron-positron data. As in Ref.~\cite{Plehn:2019jeo}, we define a sub-set of the free fit parameters for each channel and vary their mean values within the uncertainty provided by our \verb!IMinuit!~\cite{hans_dembinski_2021_5561211} fit or as stated in the papers. For more details about the uncertainty estimates for the individual channels, we refer to Ref.~\cite{Plehn:2019jeo}. We obtain envelopes around the mean values for the $e^+e^-$ cross-section data and propagate those parameters to calculate the enveloping curves of the hadronic widths and related quantities. 
 
\begin{figure}[h]
\begin{center}
\includegraphics[width=0.8\textwidth]{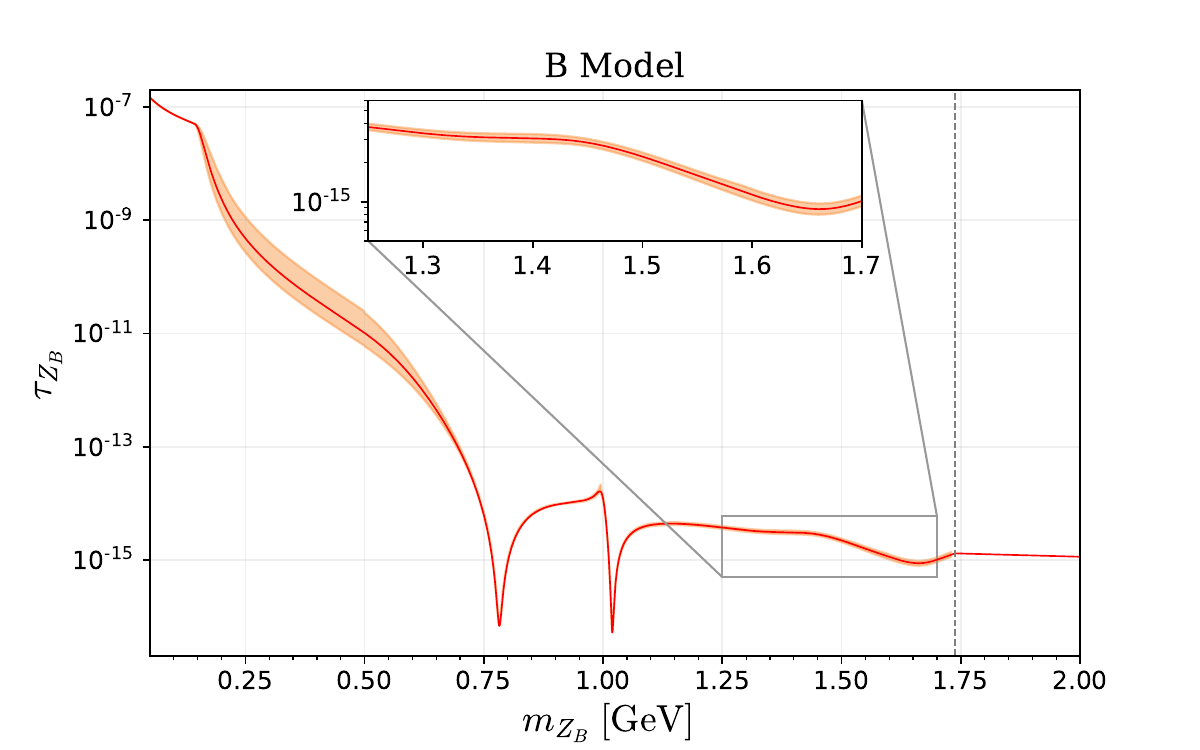}
\end{center}
\vglue -0.8 cm
\caption{\label{fig:lifetime} Uncertainty on the $Z_B$ mediator lifetime obtained by propagating the cross-section fit envelopes into the hadronic width computation. The red curve is the $Z_B$ lifetime evaluated by considering the best fit parameters for each hadronic channel width, while the orange region represents the envelope lifetime uncertainty estimate. The large uncertainties below $\sim 0.6 \; \GeV$ are caused by the lack of $\pi\gamma$ experimental data in this region. For the regions where data is available, the uncertainties always stay below the 10\% level, as we can see in the zoomed in plot. We considered $g_B = 10^{-4}$ for the lifetime calculation,  we remark, however, that the mediator coupling does not affect the uncertainties. The vertical dashed grey line indicates the hadron-quark transition.}
\end{figure}

In Fig.~\ref{fig:lifetime} we show in which way this affects the $Z_B$ mediator lifetime. As we can see, below the pion threshold, the $Z_B$ mediator decays into leptons, which can be calculated perturbatively and, hence, no error bars are included. In the mass region just above the pion threshold up to around 600 MeV the only channel present is $\mathcal{H}=\pi \gamma$. The large uncertainties in this region are justified since no data is available in this mass range as seen in Fig.~\ref{fig:Rpigamma} of appendix \ref{appx:newfit}. In an obvious way, our data-driven estimates could be, therefore, improved if new data were available below 600 MeV, in particular for the $\pi\gamma$ channel as it is the dominant hadronic channel in this region for $B$-coupled models. Around and above the $\rho, \omega$ and $\phi$ resonances, the uncertainties lie below the 10\% level. The uncertainties will be even smaller for other quantities like the branching ratios as they would affect both nominator and denominator of the ratio. Furthermore, we choose the $B$ model as an example since errors would be, if at all, mostly visible for models that do not couple to the precisely measured $2\pi$ and $4\pi$ currents with small uncertainties as well as to leptons, which would dominate the lifetime computation for masses away from the resonance peaks. Since the theoretical uncertainties are already well below the 10\% level for most of the mass range for the lifetime of this model, we refrain from further including them for all other quantities presented in the course of this paper.

%

\section{Results and Impact on Present and Future Bounds}
\label{sec:results}

We will start this section by presenting the changes in the hadronic decay widths and  branching ratios that result from  
our better assessment of the $Z_{Q}$ decays to light hadrons. 
After that we will show the consequences on present limits and future experimental sensitivities for a few models.

\subsection{Hadronic Decay Widths}
\label{sec:decwidths}

\begin{figure}[h!]
\begin{center}
\includegraphics[width=1.0\textwidth]{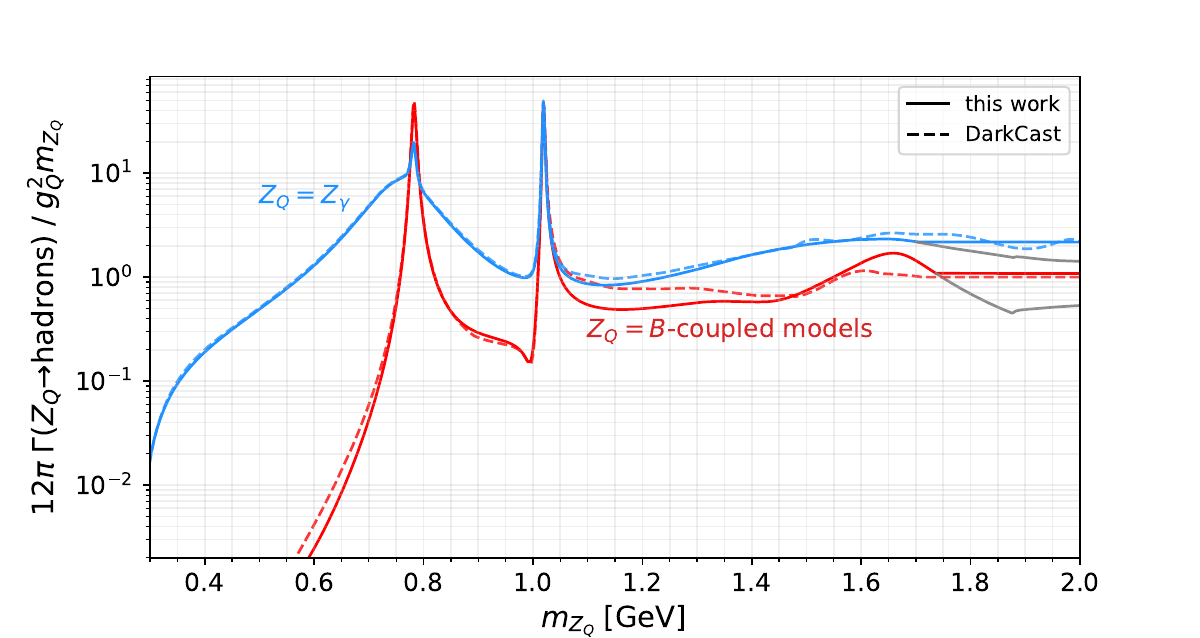}
\end{center}
\caption{\label{fig:width} Comparison of the total hadronic width  (solid lines) for the {\it dark photon} and $B$-coupled (baryophilic) $Z_Q$ models with the ones implemented in \textsc{DarkCast} (dashed lines). Around $m_{Z_Q} \approx 1.7$ GeV we make the transition to 
perturbative QCD (see discussion in section~\ref{sec:HadCalc}{\it e}).}
\end{figure}

In figure~\ref{fig:width} we show the total hadronic decay width, normalized to $g_{Q}^2 \, m_{Z_{Q}}$,  as a function of $m_{Z_{Q}}$ for the {\it dark photon} (solid blue line, $Z_{Q}=Z_\gamma$) and for all the $U(1)_{Q}$ models we discuss in this paper (solid red line). We also show for comparison the results of the previous calculation (dashed lines). The differences between the solid and dashed curves are more sizable in the region $1 \lesssim m_{Z_{Q}}/{\rm GeV} \lesssim 1.7$, where we included several new hadronic channels. Close to $1.7 \; \rm GeV$ we perform the transition to the perturbative width, which we indicate by splitting the solid curve into another grey curve that represents the hadronic width continuation. 

In the region above $1.7 \; \rm GeV$, one can see that, for the {\it dark photon} case, the width from \textsc{DarkCast} has different features in comparison with the straight perturbative line of our approach. The reason for that is related to the fact that, due to the inclusion of a small number of hadronic channels in \cite{Ilten:2018crw}, the authors considered the following strategy to reach the total $R_\mu^{\rm SM}$ curve: they take their $\gamma$-like curve to be the PDG curve above $1.48 \; \GeV$ and their calculation below this energy. Then, they define their $\rho$-like curve to be described by the $2\pi$ and $4\pi$ channels below $1.1 \; \GeV$ and to be the $\gamma$-like curve, with the $\omega$ and $\phi$ contributions subtracted, above it.

On the one hand, the method described above allowed their $\gamma$-like curve to match the  $R_\mu^{\rm SM}$ experimental calculation. On the other hand, this approximation makes the wrong assumption that all the other neglected hadronic channels contribute as $\rho$ components. As a result, we can see from the figure that right before the transition the red solid line is larger than the dashed one, since baryophilic models do not couple to the $\rho$ current, which means for \textsc{DarkCast} that all the other possible hadronic channels that they did not consider will not couple to the $Z_Q$ bosons of these models. We can also see that, for the case of the $B$-coupled models, the dashed line becomes a straight line close to the transition. This behavior is a consequence of the $\omega$ and $\phi$ contributions, that also transition to perturbative values close to $1.6 \; \GeV$ and $1.7 \; \GeV$, respectively. However, the red solid line establishes a little bit above the dashed one due to our inclusion of QCD corrections in eq.~(\ref{eq:Rpert}). 

Another aspect that is important to highlight is the difference for low energies. The two red lines differ close to $0.6 \; \GeV$ as a result of the divergences in the calculation of the $\pi \gamma$ channel, as explained in the previous section. This specific channel has a great impact because is the first hadronic channel that couple to the baryophilic model currents. As we will see next, the branching ratios will also modify as a consequence of the above mention disparities.

\subsection{Branching Ratios}
\label{sec:brs}
\begin{figure}[h]
\begin{center}
\includegraphics[width=1.0\textwidth]{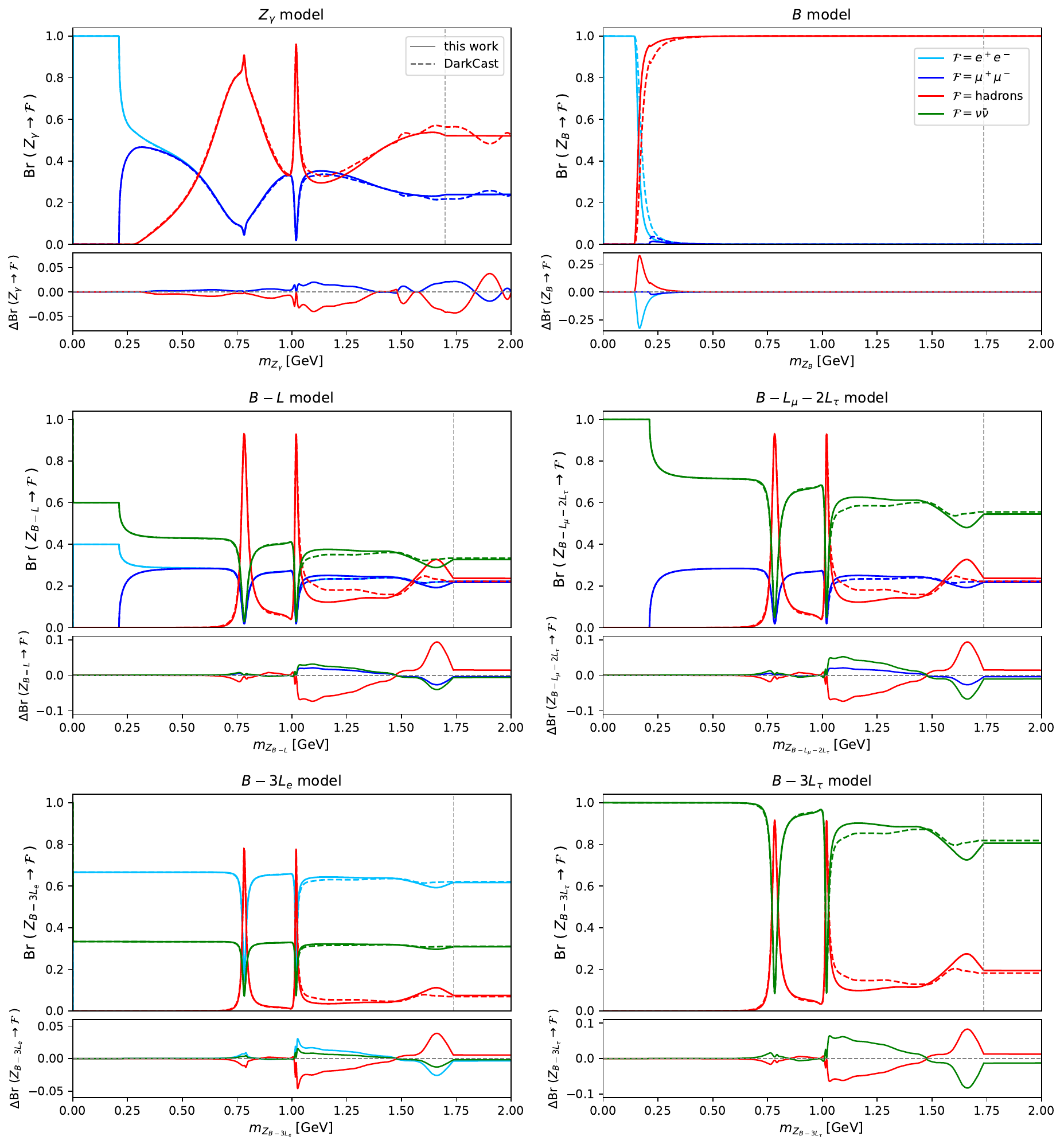}
\end{center}
\vglue -0.8 cm
\caption{\label{fig:bfrac} Comparison of the leptonic and hadronic branching ratios (solid lines) with the ones from \textsc{DarkCast} (dashed lines) for some chosen models. The vertical dashed gray line indicates the transition from non-perturbative to perturbative calculations as described in the text. In the lower panel of each figure we show the deviation $\Delta \rm Br$, i.e. our branching ratio minus the \textsc{DarkCast} one.}
\end{figure}

Now we examine how differences in the hadronic channels affect the 
branching ratios of the models of interest. In figure~\ref{fig:bfrac} in the top panel of each model, we 
show the branching ratios into $e^+e^-$ (light blue), $\mu^+\mu^-$ (blue), neutrinos (green) and hadrons (red) as a function of the mass of the vector boson.  The solid (dashed) lines represent the results of  our (previous) calculations. In the bottom panel
of each plot we show the branching ratio difference between the two calculations.

\begin{figure}[h]
\begin{center}
\includegraphics[width=1.0\textwidth]{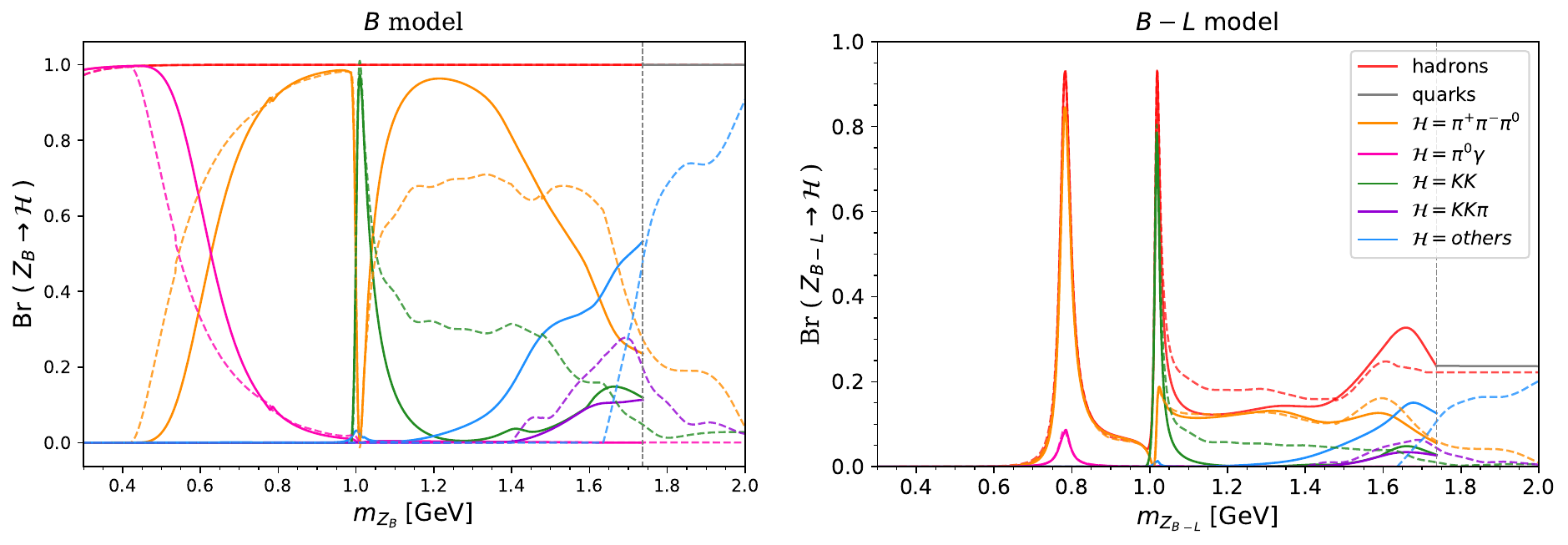}
\end{center}
\vglue -0.8 cm
\caption{\label{fig:bfracHad} Comparison of the individual contributions to the total hadronic branching ratio between our calculations (solid lines) and  \textsc{DarkCast} (dashed lines) for the $B$ (left panel) and $B-L$ (right panel) models. The individual branching ratios for the other $B$-coupled models behave in a similar way to the $B-L$ model. The vertical dashed gray line indicates the transition from non-perturbative to perturbative calculations as described in the text.}
\end{figure}

We show, for reference, the $Z_\gamma$ case as well as the pure $Z_B$. In the $Z_\gamma$ case, \textsc{DarkCast} predicts a larger branching ratio into hadrons than 
us in the range $0.25 \lesssim m_{Z_{\gamma}}/{\rm GeV} \lesssim 1.8$, but the difference is always less than 5\%. The discrepancy 
between the two calculations for $Z_B$ is, on the other hand, 
more visible for $0.2 \lesssim m_{Z_{B}}/{\rm GeV} \lesssim 0.4$
because the previous calculation underestimates the 
$\pi^0 \gamma$ contribution (see section~\ref{sec:HadCalc} {\it b}). In this region, the difference can be as large as $\sim $ 30\%. In spite of the fact that for larger values of $m_{Z_{B}}$ the hadronic branching ratios seem to coincide, we see in the left panel of figure~\ref{fig:bfracHad} that the contributions of each hadronic mode is quite different. For instance, our calculation predicts a much smaller (larger) contribution of the $KK$ (3$\pi$) final state in the region $1.0 \lesssim m_{Z_{B}}/{\rm GeV} \lesssim 1.5$.

For the $B-L$, $B-L_\mu-2L_\tau$, $B-3L_e$ and $B-3L_\tau$ models~\footnote{We do not show here the branching ratios for the models $B-3L_\mu$ and $B-L_e-2L_\tau$, 
because they are similar to $B-3L_e$ and  $B-L_\mu-2L_\tau$, respectively. One only has to 
exchange  the lines $\mathcal{F}= e^+e^- \leftrightarrow \mathcal{F}= \mu^+\mu^-$.}, the hadronic contribution to the branching ratio in the 
region $1.0 \lesssim m_{Z_{Q}}/{\rm GeV} \lesssim 1.75$ is 
 sometimes overestimated  (due to $KK$ mode) sometimes underestimated (due to higher resonances) by \textsc{DarkCast}, generally influencing the charged lepton and neutrino decay contributions by a few to almost 10\% for some values of $m_{Z_Q}$ in some of the models.
We illustrate these changes  in the contributions of the hadronic final states for these models showing them explicitly for the $B-L$ model in the right panel of figure~\ref{fig:bfracHad}.


\subsection{Repercussions on Current Limits and Future Sensitivities}
\label{sec:limits}
To discuss the effect of our reevaluation of the light hadron contributions to $Z_{Q}$ decays on experimental limits for these models in the range $100 \, {\rm MeV} \leq m_{Z_Q}\leq 2\, {\rm GeV}$, we have implemented the results of our  calculations in the \textsc{DarkCast} and \textsc{FORESEE} codes. \textsc{DarkCast} is a code that recasts experimental limits on {\it dark photon} searches to obtain limits on vector boson mediators with couplings to SM fermions. See Ref.~\cite{Ilten:2018crw} for more details on the recasting procedure for the different types of experimental data we have used to obtain the limits presented here. 
\textsc{FORESEE} (FORward Experiment SEnsitivity Estimator) is a package 
that can be used to calculate the expected sensitivity for BSM physics 
of future experiments placed in the forward direction far from the proton-proton interaction point at the LHC. See Ref.~\cite{Kling:2021fwx} for more information on the code.

\subsubsection{Current Experimental Limits}

To obtain the exclusion regions in the $g_Q \times m_{Z_Q}$ plane 
for the various models of interest, we consider the following experimental searches:

\begin{enumerate}
\item $Z_Q$ produced in the electron fixed target experiments APEX~\cite{APEX:2011dww} and A1~\cite{A1:2011yso,Merkel:2014avp}
by Bremsstrahlung followed by the prompt decay  $Z_Q \to e^+ e^-$~\cite{NA64:2016oww,NA64:2017vtt,Banerjee:2019pds};
and in NA64 followed by the prompt decay  $Z_Q \to$ invisible ($\nu \bar \nu$);
\item $Z_Q$ produced  via $\pi^0 \to\gamma Z_Q$ in the proton beam dump experiments 
LSND~\cite{LSND:1997vqj,Bauer:2018onh}, 
PS191~\cite{Bernardi:1985ny} and  NuCal~\cite{Blumlein:1991xh} 
as well as via $\eta \to\gamma Z_Q$ in CHARM~\cite{CHARM:1985anb} and via proton Bremsstrahlung in NuCal~\cite{Blumlein:1990ay}, all of them followed by $Z_Q \to e^+ e^-$;
\item $Z_Q$ produced in the electron beam dump experiment E137  followed by $Z_Q \to e^+ e^-$~\cite{Bjorken:2009mm,Andreas:2012mt}; 
\item $Z_Q$ produced by radiative return in the $e^+e^-$ annihilation experiments BESIII, BaBar and KLOE or by muon Breemstrahlung in Belle-II. In BaBar, one searches for the decay modes $Z_Q \to e^+e^-,\, \mu^+\mu^-$~\cite{BaBar:2014zli}
and $Z_Q \to$ invisible ($\nu \bar \nu$)~\cite{BaBar:2017tiz}, 
in BESSIII for the decay modes $Z_Q \to e^+e^-,\, \mu^+\mu^-$~\cite{BESIII:2017fwv}
and in KLOE for the decay modes $Z_Q \to e^+e^-$~\cite{Anastasi:2015qla} and
$Z_Q \to \mu^+\mu^-$~\cite{KLOE-2:2014qxg,KLOE-2:2018kqf}. For KLOE we also use data for the search $\phi \to \eta Z_Q\, , Z_Q \to e^+e^-$~\cite{KLOE-2:2012lii}. In Belle-II, one searches for $Z_Q \to$ invisible ($\nu \bar \nu$)~\cite{Belle-II:2019qfb};
\item $Z_Q$ produced in pp collisions at the LHCb experiment either by meson decays or the Drell-Yan mechanism with the subsequent displaced or prompt decay $Z_Q \to \mu^+ \mu^-$~\cite{LHCb:2017trq,LHCb:2019vmc};
\item $Z_Q$ produced in  kaon decay experiments via $\pi^0 \to \gamma Z_Q$ 
followed by the prompt decay $Z_Q \to e^+e^-$ at  NA48/2~\cite{NA482:2015wmo} or by $Z_Q \to$ invisible ($\nu \bar \nu$) at NA62~\cite{NA62:2019meo}.

\end{enumerate}

\begin{figure}[htb]
\begin{center}
\includegraphics[width=1.0\textwidth]{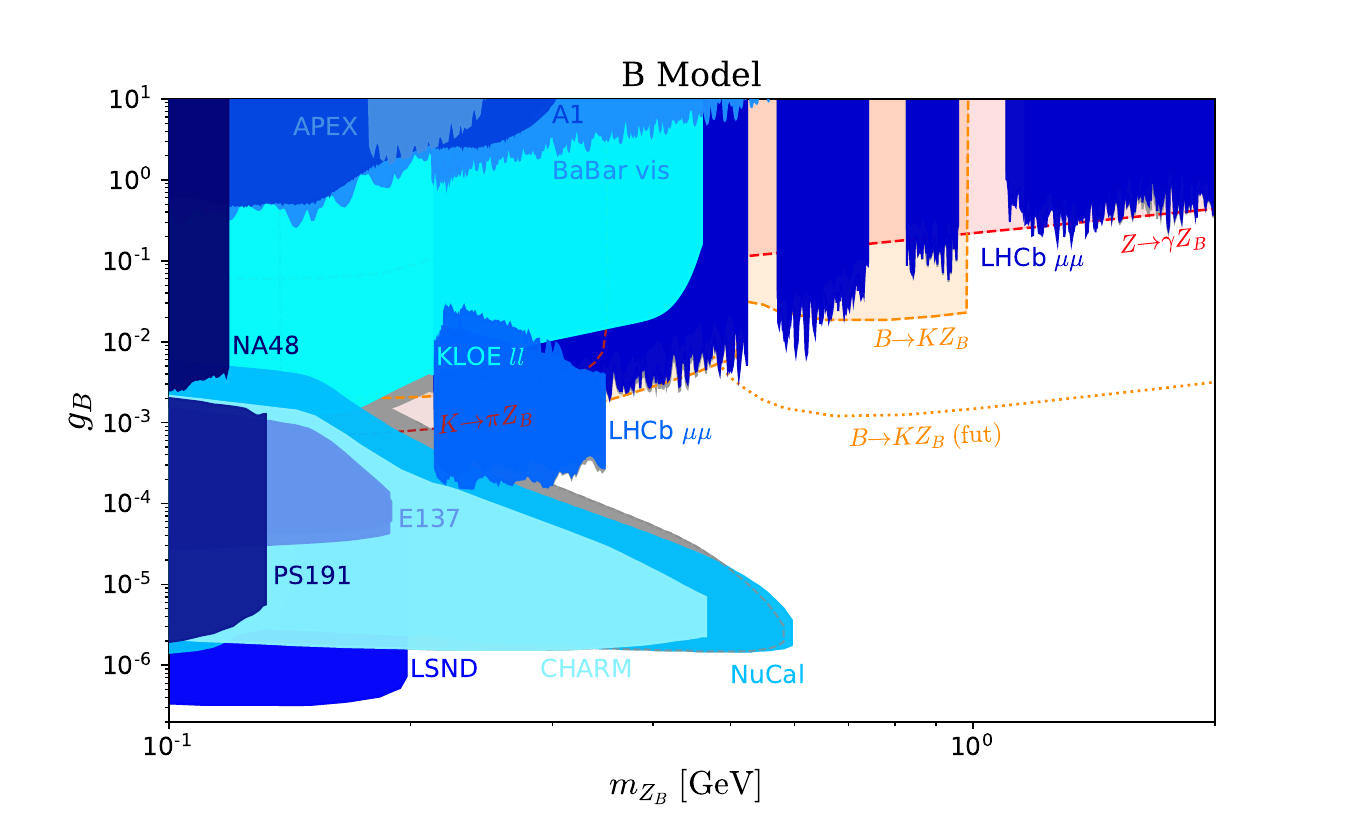}
\end{center}
\vglue -0.8cm
\caption{\label{fig:BBlim}  In blue the excluded regions in the plane $g_B \times m_{Z_B}$ we obtained using data from the electron Bremsstrahlung experiments APEX~\cite{APEX:2011dww} and A1~\cite{A1:2011yso,Merkel:2014avp}, the proton beam dump experiments PS191~\cite{Bernardi:1985ny}, NuCal~\cite{Blumlein:1990ay,Blumlein:1991xh} and CHARM~\cite{CHARM:1985anb}, the electron beam dump experiment E137~\cite{Bjorken:2009mm,Andreas:2012mt}, the $e^+e^-$ annihilation experiments BaBar~\cite{BaBar:2014zli} and KLOE~\cite{Anastasi:2015qla,KLOE-2:2012lii}, the LHCb experiment~\cite{LHCb:2017trq,LHCb:2019vmc}, NA48~\cite{NA482:2015wmo} and LSND~\cite{LSND:1997vqj,Bauer:2018onh}. 
In gray the region excluded by the previous calculation~\cite{Ilten:2018crw}, but still allowed by this work. We also show the limits from $B \to K Z_B$, $K^\pm \to \pi^\pm Z_B$ and $Z\to \gamma Z_B$ taken from~\cite{Dror:2017ehi} for completeness, where the dashed lines represent current bounds and the dotted lines future predictions.}
\end{figure}

We start by presenting the differences on the limits for the 
$U(1)_B$ model as it highlights the consequences of the improvements of our calculations. In all the plots in blue (green) we show the 
exclusion regions for $Z_{Q}$ decaying to $e^+e^-$ and $\mu^+ \mu^-$ pairs (neutrinos).
In figure~\ref{fig:BBlim} we show in blue the recasted limits for various experiments using our  calculations. In gray we can see an extra region that would be excluded by \textsc{DarkCast}, but not by this work. This is particularly visible for $0.2 \lesssim m_{Z_{B}}/{\rm GeV} \lesssim 0.4$ where the underestimation of the $\pi^0 \gamma$ contribution in the previous calculation yields to an enhanced $Z_{B}\to e^+e^-$ signal prediction. There are also regions where our calculation results in an increase of the exclusion bounds. For instance, we show in figure~\ref{fig:BBlim} in gray the contour for the NuCal limits obtained with \textsc{DarkCast}. As one can see, there is a region previously allowed that we can exclude now. This effect is also a consequence of the difference in the lifetime calculation, that is more prominent for the $B$ model, and has a deep impact specially for beam-dump experiments.

There are, however, two caveats here. The first is the fact that the model is anomalous. As it has been shown in Ref.~\cite{Dror:2017ehi,Dror:2017nsg} light vectors coupled to SM particles and non-conserved currents enhance the rate of meson decays such as $B \to K Z_{B}$ and $K^\pm \to \pi^\pm Z_B$ as well as the Z boson decay $Z \to \gamma Z_{B}$. Those limits mostly lie in areas that have been covered by LHCb with the exception of filling unconstrained areas in the vector meson resonance region. Furthermore, the future $B\to K Z_B$ prediction is expected to cover a sizable part of the region $0.5~{\rm GeV} \lesssim m_{Z_B}$.
The second is related to the coupling to
leptons, as for all experimental limits the light vector boson is supposed to decay to  $e^+e^-$ and/or $\mu^+ \mu^-$ (BaBar and LHCb).
 Although $Z_{B}$ does not couple directly to charged leptons, 
there is a one-loop induced kinetic mixing between $Z_B$ and the photon~\cite{Carone:1994aa}. However, the magnitude of this coupling will depends on the choice of the renormalization scale so it cannot 
be determined unambiguously. In the \textsc{DarkCast} code, which we use, it is taken to be simply $e g_B/(4\pi)^2$, so the limits involving this coupling to charged leptons have to be regarded with caution.

\begin{figure}[htb]
\begin{center}
\includegraphics[width=1.0\textwidth]{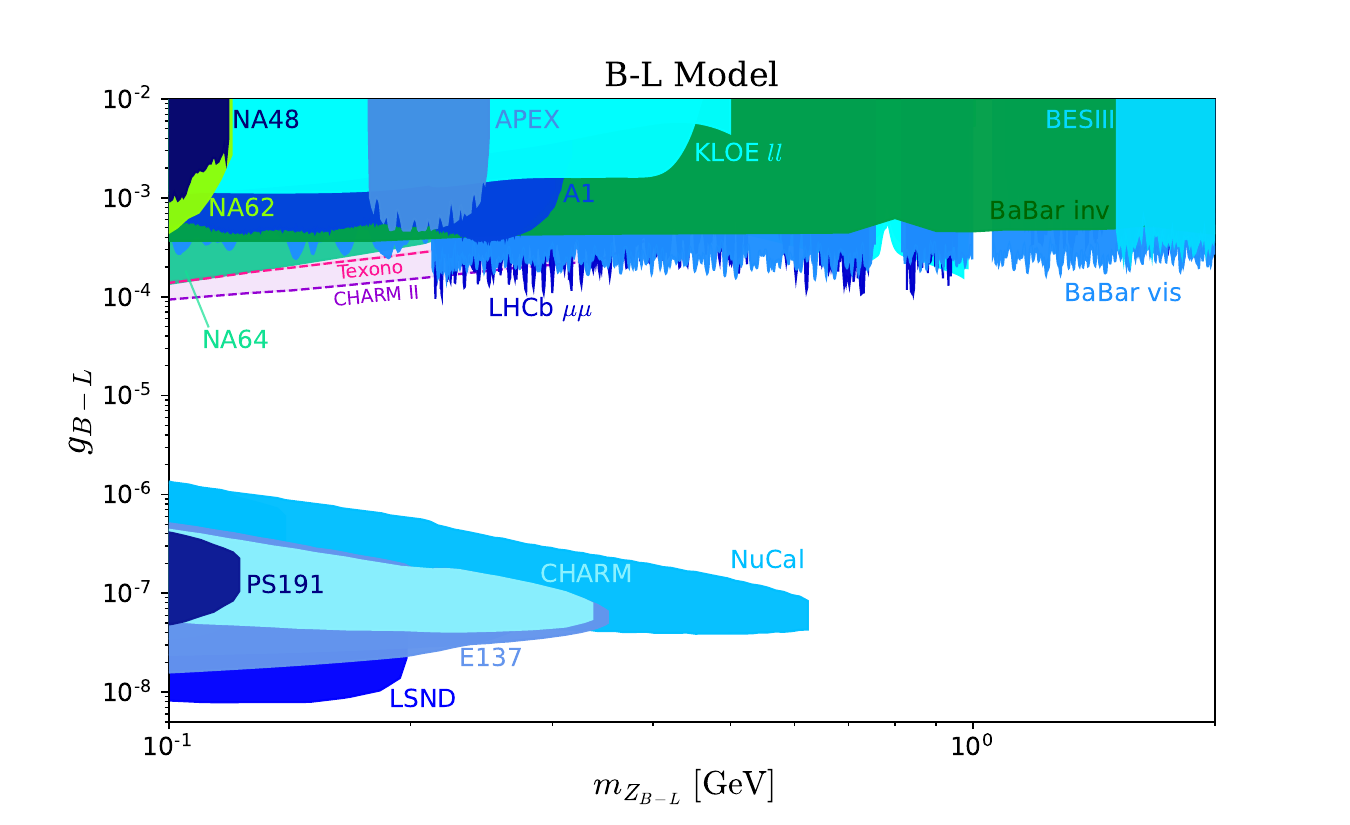}
\end{center}
\vglue -0.8 cm
\caption{\label{fig:BLlim}  Same as figure~\ref{fig:BBlim} but for the $B-L$ model. In blue (green) the excluded regions for $Z_{B-L}$ decaying to charged lepton ($\nu \bar \nu$) pairs.
Beside the data already included in  figure~\ref{fig:BBlim}, here we also include data from KLOE in the $\mu^+\mu^-$ final state~\cite{KLOE-2:2014qxg,KLOE-2:2018kqf}, from BaBar~\cite{BaBar:2017tiz},
NA62~\cite{NA62:2019meo} and NA64~\cite{NA64:2016oww,NA64:2017vtt,Banerjee:2019pds} invisible searches, and from BESIII~\cite{BESIII:2017fwv}. We also show with dashed lines the limits from the neutrino experiments Texono \cite{TEXONO:2009knm} (red) and CHARM-II \cite{GEIREGAT1990271} (purple) that were taken from  \cite{Bauer:2018onh}. }
\end{figure}
\begin{figure}[hbt]
\vglue -1.5 cm
\begin{center}
\includegraphics[width=1.0\textwidth]{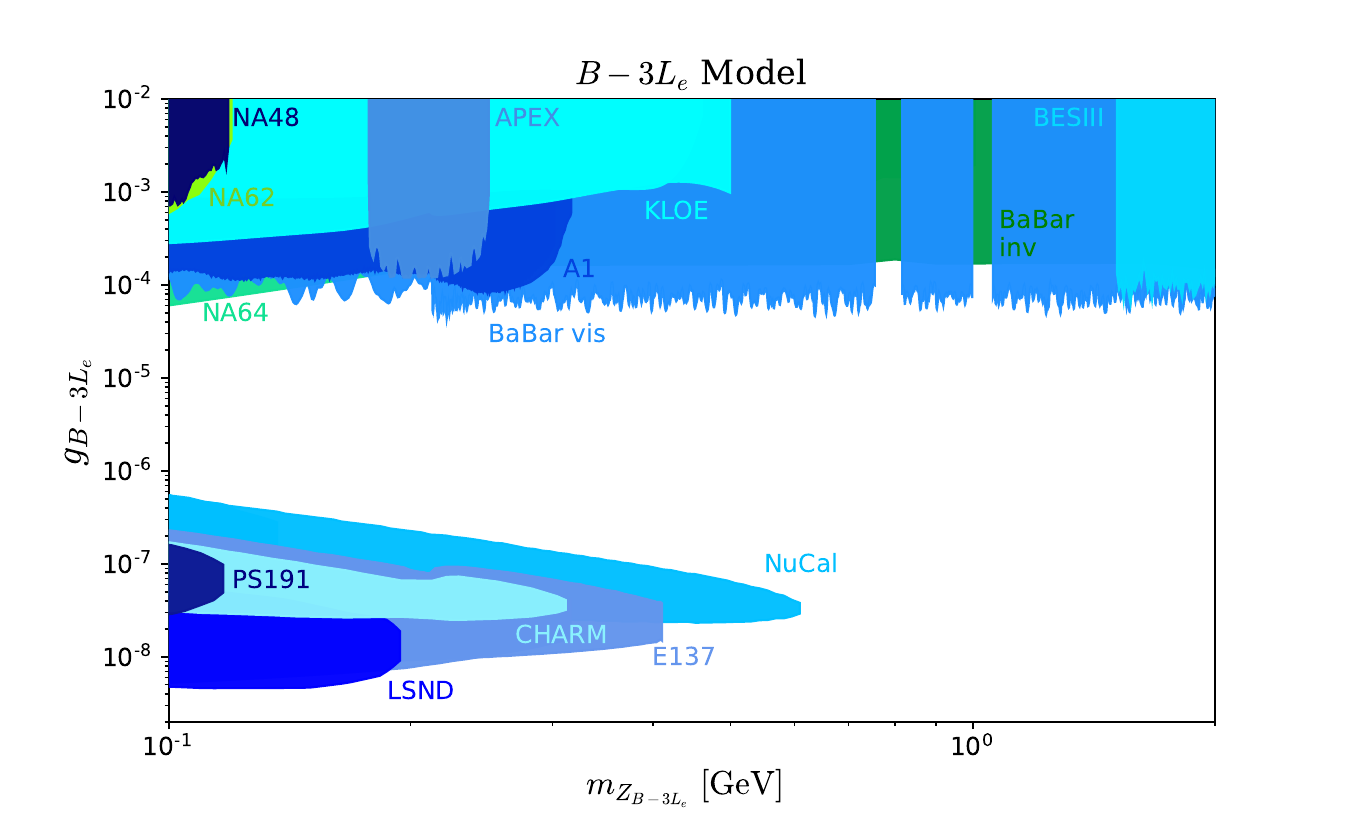}
\end{center}
\vglue -0.8 cm
\caption{\label{fig:B3Lelim}Similar to figure~\ref{fig:BLlim} but for the $B-3L_e$ model. Here there is no contribution from the LHCb experiment or from KLOE due to the absence of muon couplings.}
\end{figure}
\begin{figure}[htb]
\vglue -.59 cm
\begin{center}
\includegraphics[width=1.0\textwidth]{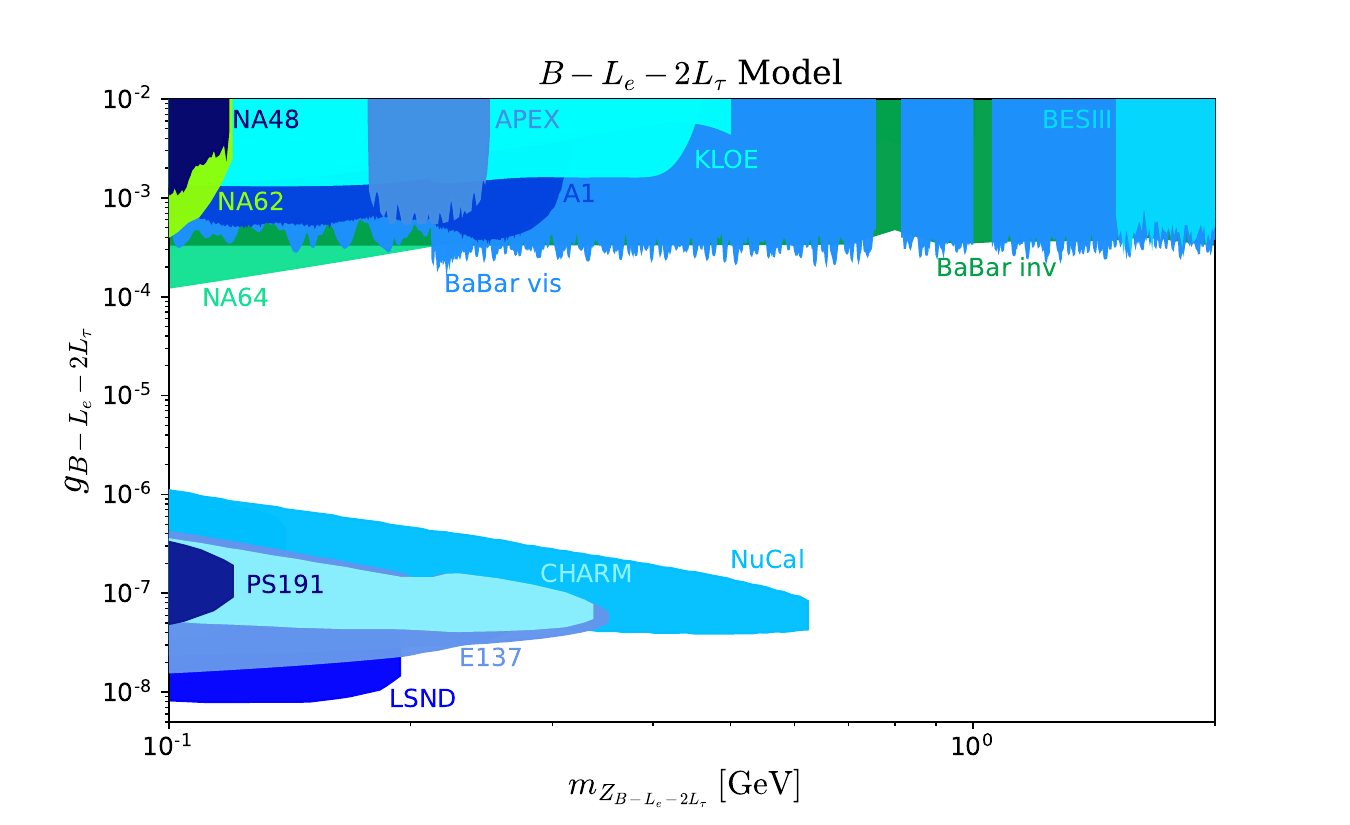}
\end{center}
\vglue -0.8 cm
\caption{\label{fig:BLe2Ltaulim}Similar to figure~\ref{fig:B3Lelim} but for the $B-L_e-2L_\tau$ model.}
\end{figure}

Next we show the exclusion regions for some of the models we have considered. Although in the case of current limits, the differences caused by our calculations are not very visible in the combined plot, they will affect the sensitivity of future experiments as we will see shortly. In figure~\ref{fig:BLlim} we show the exclusion region for $Z_{B-L}$ in the plane $g_{B-L} \times m_{Z_{B-L}}$. Here the differences are small as they practically do not affect $Z_{B-L} \to e^+e^-,\,  \mu^+ \mu^-$ and $\bar{\nu}\nu$.  However, since this model is of great interest and  we have some recent data from LHCb, NA62 and NA64, we decided to present here. We also include for completeness the limits from the neutrino experiments Texono \cite{TEXONO:2009knm,Lindner:2018kjo,Bilmis:2015lja} and CHARM-II \cite{GEIREGAT1990271,Lindner:2018kjo,Bilmis:2015lja}  that were taken from \cite{Bauer:2018onh}. These limits do not depend on leptonic decays and therefore are independent of the hadronic branching ratios. We do not show the limit from the Borexino \cite{Bellini:2011rx,Harnik:2012ni,Amaral:2020tga} neutrino experiment since the NA64 and CHARM-II limits cover it in the mass range considered in this study. In figure~\ref{fig:B3Lelim} we show the exclusion region for $Z_{B-3L_e}$ which is similar but does not contain the constraints from LHCb, and KLOE in the $\mu^+ \mu^-$ final state.

 Finally, in figure~\ref{fig:BLe2Ltaulim} we show the limits for the  $B-L_e-2L_\tau$ model. The $B-3L_\mu$ and $B-L_\mu-2L_\tau$ models only have bounds from LHCb (prompt), NA62  and Belle-II, while the $B-3 L_\tau$ model  only has bounds from NA62. We do not show them here but refer to \cite{Bauer:2020itv} for a comprehensive analysis of $B-3L_i$ models.
 Note that all experimental searches reported here 
 look for either  leptonic or invisible (neutrino) decays of the 
 vector mediator. Hadronic decays, however, especially close to the 
 vector resonances, could in general be probed.

\newpage

\subsubsection{Future Experimental Sensitivities}

Here we discuss how our better assessment of the $Z_{Q}$ decay to light hadrons can affect the sensitivity of various high intensity frontier experiments that can probe them in the near future. 

The ForwArd Search ExpeRiment (FASER) is a relatively small cylindrical detector located along the LHC beam axis at approximately 480 m downstream of the ATLAS detector interaction point. The aim is to search for long lived particles profiting of the luminosity and boost of the LHC beam.
There are two proposed phases for FASER. In the first phase, named FASER, the detector will be 1.5 m long with a diameter of 20 cm and will operate from 2022 to 2024~\cite{FASER:2021ljd}, being exposed to an expected integrated luminosity of 150 fb$^{-1}$~\cite{FASER:2018eoc}. In the second phase, named FASER 2, the detector will be 5 m long with a diameter of 2 m and is expected to take data in the high luminosity LHC era, being exposed to an integrated luminosity of 3 ab$^{-1}$. {\it Dark photons} can be produced by meson decays, $pp \to Z_{\gamma} pp$ (Bremsstrahlung) as well as by direct production in hard scattering. It is important to highlight that, in contrast to the majority of current experimental searches, that rely on leptonic decay signals, the FASER detector will also be sensitive to hadronic final states. Hence, it is crucial to provide a correct hadronic description in order to precisely compute the experiment expected sensitivity.

\begin{figure}[h]
\begin{center}
\includegraphics[width=0.49\textwidth]{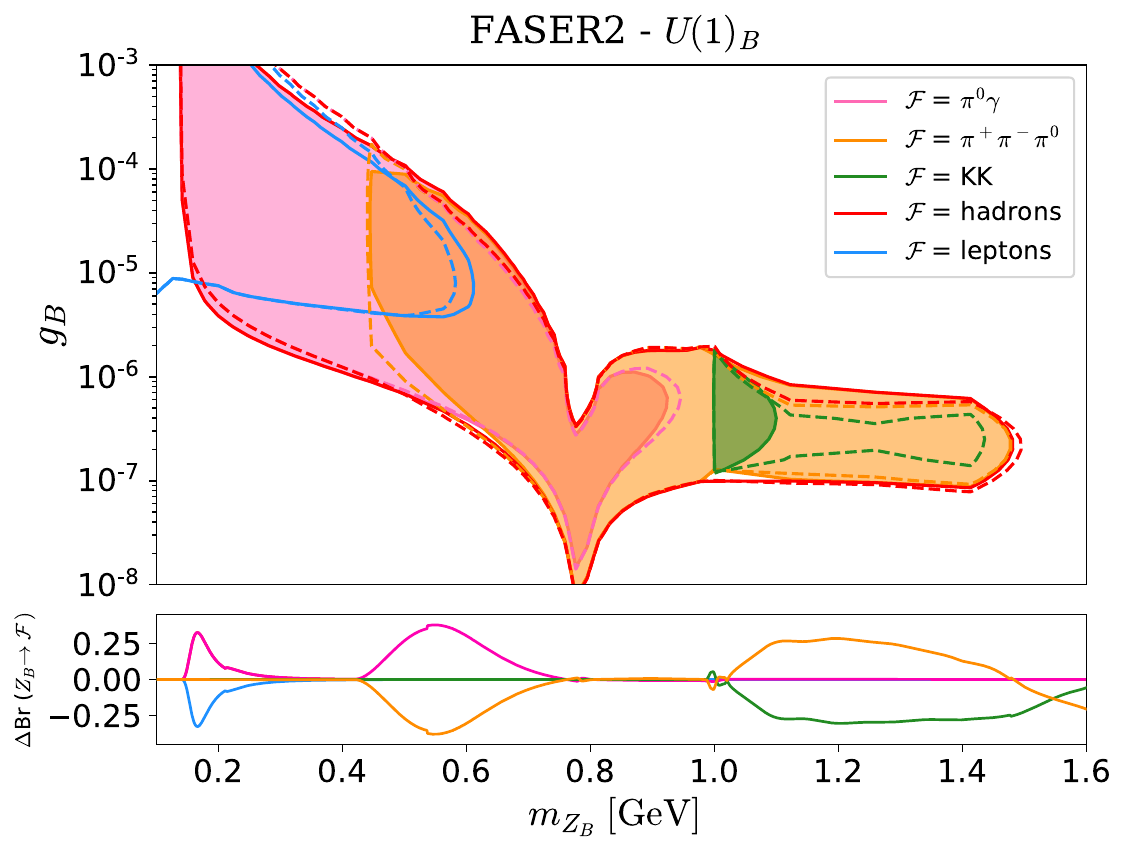}
\includegraphics[width=0.48\textwidth]{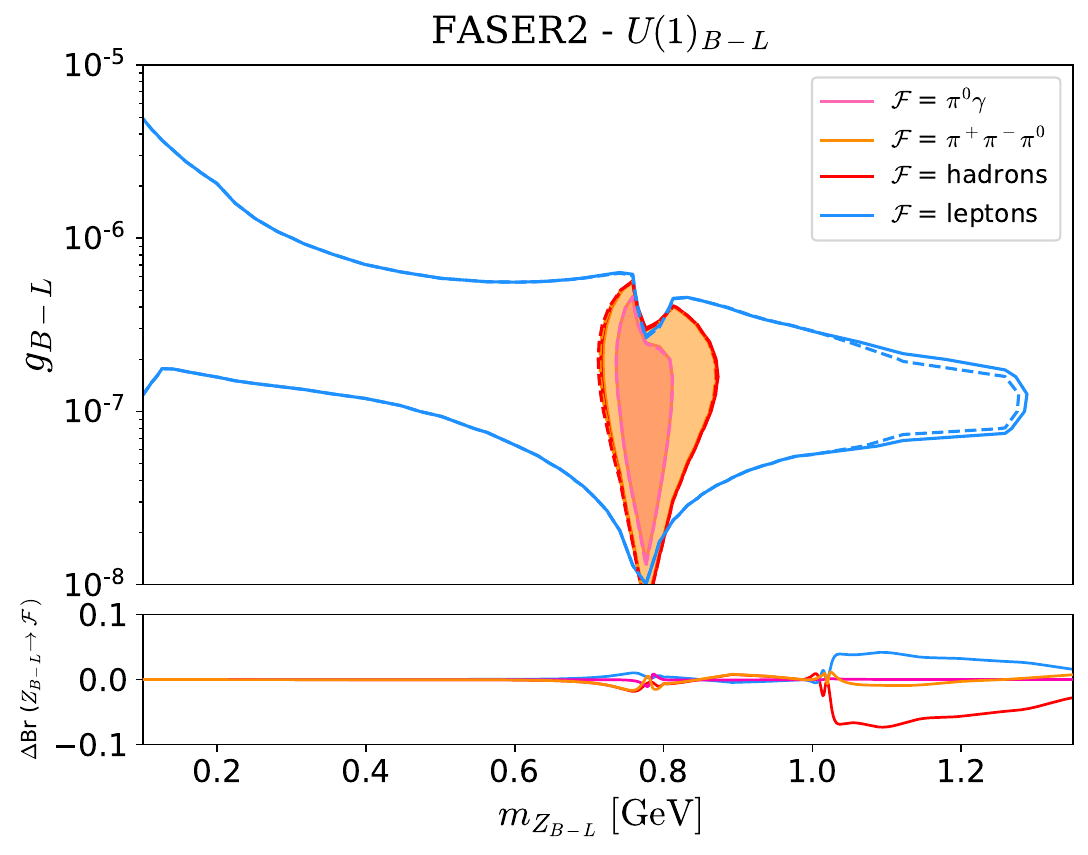}
\end{center}
\vglue -0.8cm
\caption{\label{fig:reachB} 
Expected sensitivity for the $B$ (left panel) and $B-L$ (right panel) models for FASER2 using our calculations for the branching fractions 
implemented in the \textsc{FORESEE} code.
The various final state contributions  are highlighted by different colors 
as in figure~\ref{fig:bfracHad}: $\pi \gamma$ (pink), $3 \pi$ (orange), $KK$ (green) and leptons (blue).
The dashed lines, using the same color scheme, show the results using the \textsc{FORESEE} code and \textsc{DarkCast} branching ratios. In the bottom panels we also show for each model the difference of the branching ratio between our calculation and 
\textsc{DarkCast}.}
\end{figure}

In figure~\ref{fig:reachB} we show the sensitivity  for the  $B$ (left panel) and $B-L$ (right panel) models expected for FASER 2 using the \textsc{FORESEE} code~\cite{Kling:2021fwx} with 
the implementation of the branching fractions we have calculated.
We highlight on these figures the various final state signal contributions  
by using different colors: $\pi \gamma$ (pink), $3 \pi$ (orange), $KK$ (green) and leptons (blue). The dashed lines using the same color scheme are the \textsc{DarkCast} predictions for each mode. 
We also show in the lower part of these plots the difference of the branching ratio between our calculation and \textsc{DarkCast}. Here we can appreciate that although the final sensitive regions do not differ very much from the one predicted by the previous calculation, the contributions from the different final states are not the same.

The proposed fixed target facility to Search for Hidden Particles (SHiP) at the CERN SPS 400 GeV proton beam~\cite{Alekhin:2015byh} is also able to search for {\it dark photons}, as well as other vector gauge bosons that couple to the gauged baryon number B, in the GeV mass range. It is expected to receive a flux of $2 \times 10^{20}$ protons on target in 5 years. The beam will hit a Molybdenum and Tungsten target, followed by a hadron stopper and by  a system of magnets to sweep muons away. The detector consists of a long decay volume that starts at about 60 m downstream from the primary target and is about 50 m long followed by a tracking system to identify the decay products of the {\it hidden particles}, for more details see \cite{Bonivento:2013jag}. At SHiP {\it dark photons} can be produced by meson decays, Bremsstrahlung and QCD. By recasting the projected constraints for the {\it dark photon} model from Bremsstrahlung production given in figure~2.6 of ref.~\cite{Alekhin:2015byh} we compute the sensitivity of other models. 

\begin{figure}[h]
\begin{center}
\includegraphics[width=0.49\textwidth]{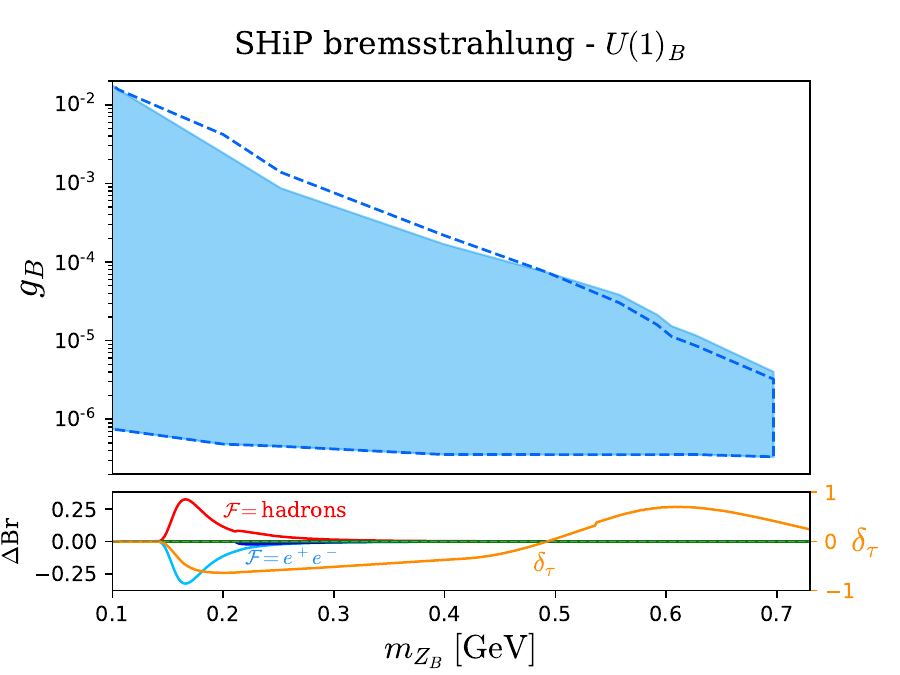}
\includegraphics[width=0.48\textwidth]{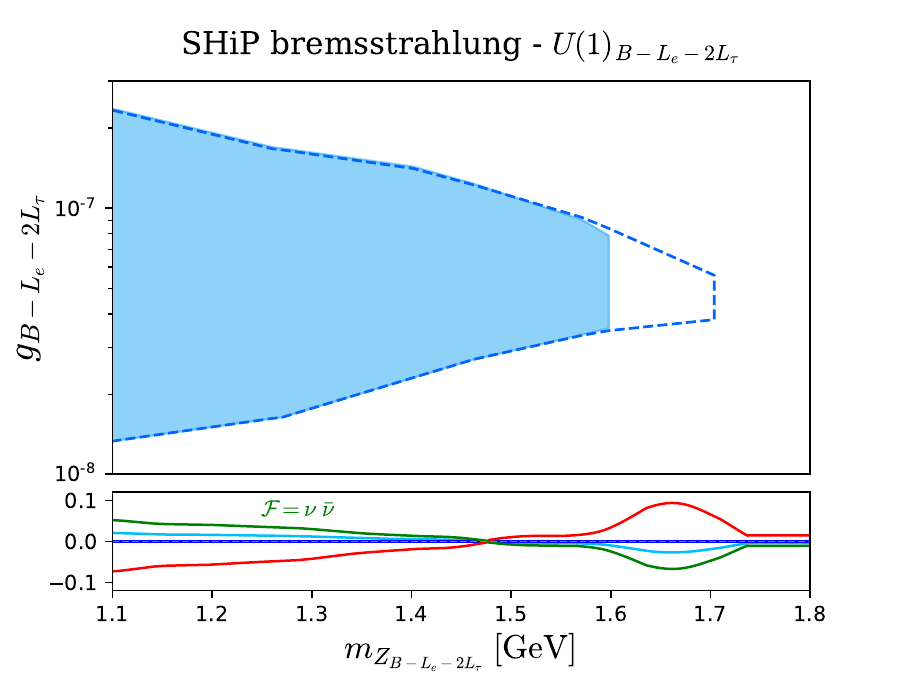}
\end{center}
\vglue -0.8cm
\caption{\label{fig:reachSHiP}  
Expected sensitivity for the $B$ (left panel) and $B-L_e-2L_\tau$ (right panel) models for SHiP Bremsstrahlung production from our calculation (solid light blue) and \textsc{DarkCast} (dashed lines).
In the bottom panels we show for each model the difference of the branching ratio between the two calculations, and for the $B$ model the corresponding difference in lifetime ($\delta_\tau$, in orange).
}
\end{figure}

In figure~\ref{fig:reachSHiP} we compare the sensitivity of the SHiP Bremsstrahlung production search for $Z_{B}$ (left panel)  and $Z_{B-L_e-2L_\tau}$ (right panel)
predicted by us (solid light blue) and \textsc{DarkCast} (dashed line).
On the bottom panels we show again the difference in the predicted branching ratios between the two calculations. For the $B$ model we also show the corresponding difference in lifetime ($\delta_\tau$, in orange).
If the lifetime is too short $Z_{B}$ will not  be able to reach the detector. So the difference with \textsc{DarkCast} comes from the smaller lifetime (for $m_{Z_B} \lesssim 0.5$ GeV) and larger lifetime (for $m_{Z_B} \gtrsim 0.5$ GeV) predicted by our calculation. In the case of the $B-L_e-2L_\tau$ model, the predicted SHiP sensitivity stops earlier, at a mass of about 1.6 GeV, due to the increase of the hadronic final state modes above this mass.

Belle-II is a high luminosity  B-factory experiment at the SuperKEKB $e^+ e^-$ collider in Japan operating at center of mass energies in the region of the $\Upsilon$ resonances. It can search for $Z_{\gamma}$ produced via the initial-state radiation (ISR) reaction $e^+e^- \to \gamma_{\rm ISR} \, Z_{\gamma}$, with $Z_{\gamma}$ decaying  to all kinetically accessible light charged states. The signature for a {\it dark photon} promptly decaying into leptons is a peak in the distribution of the reconstructed mass of the final lepton pair. We use the projected sensitivity for the visible decay modes $Z_\gamma \to e^+ e^-, \mu^+\mu^-$ from Fig.~211 of ref.~\cite{Belle-II:2018jsg}, corresponding to a total integrated luminosity of 50 ab$^{-1}$, to recast Belle-II {\it dark photon} limit to other models (visible searches). This experiment can also look for  $e^+e^- \to \gamma_{\rm ISR}\, Z_{Q}, Z_{Q} \to $ invisible, by searching for mono-energetic ISR single photons. We use their projected sensitivity for this invisible decay mode taken from figure~209 of ref.~\cite{Belle-II:2018jsg} to calculate the sensitivity of models where $Z_{Q} \to \nu \bar \nu$ can occur (invisible searches).

\begin{figure}[htb]
\begin{center}
\includegraphics[width=0.49\textwidth]{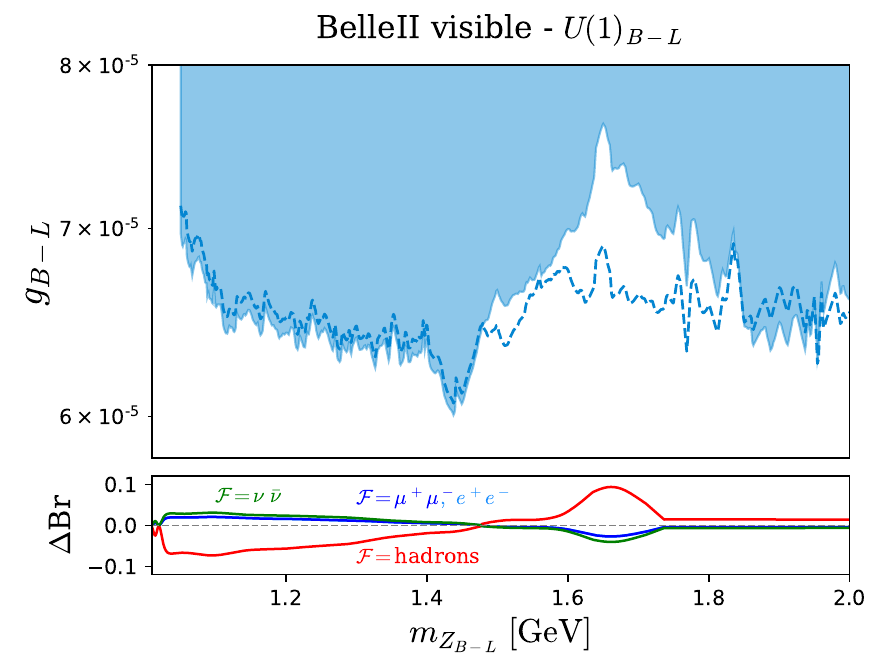}
\includegraphics[width=0.49\textwidth]{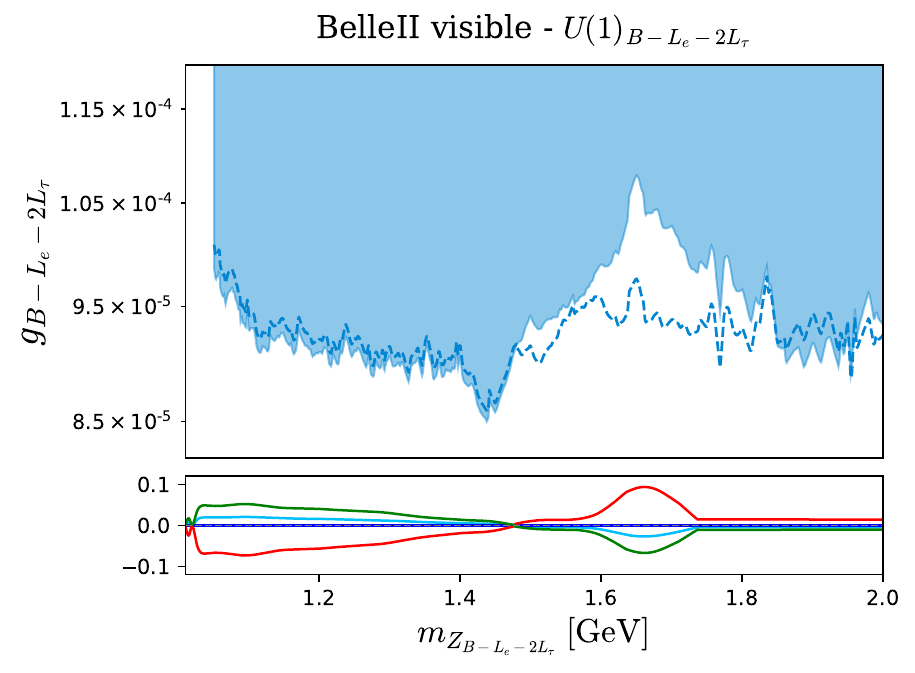}
\includegraphics[width=0.5\textwidth]{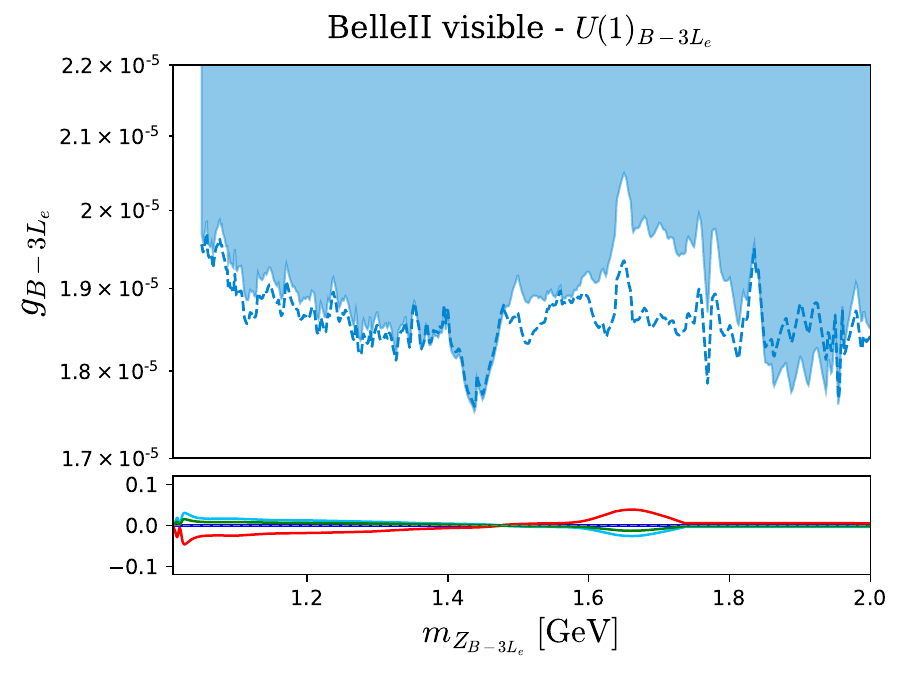}
\end{center}
\vglue -0.8cm
\caption{\label{fig:reachBelleVis} Similar to figure~\ref{fig:reachSHiP} but 
for Belle II visible searches and for the following models: $B-L$ (top left panel), $B-L_e-2L_\tau$ (top right panel) and $B-3L_e$ (bottom panel). 
In solid light blue (dashed line) we show our (\textsc{DarkCast})
results.}
\end{figure}
\begin{figure}[bht]
\begin{center}
\includegraphics[width=0.48\textwidth]{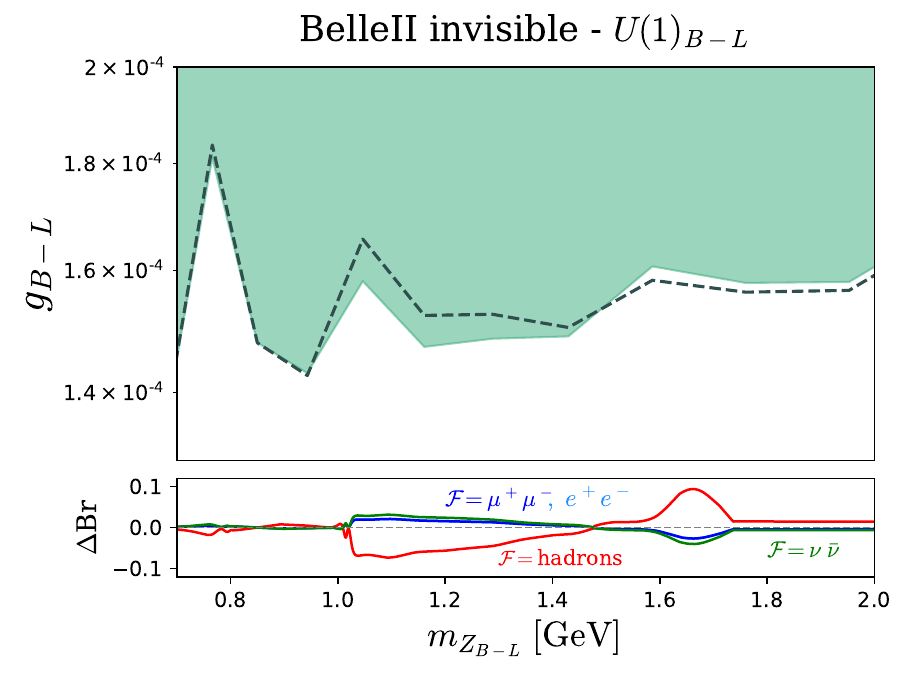}
\includegraphics[width=0.46\textwidth]{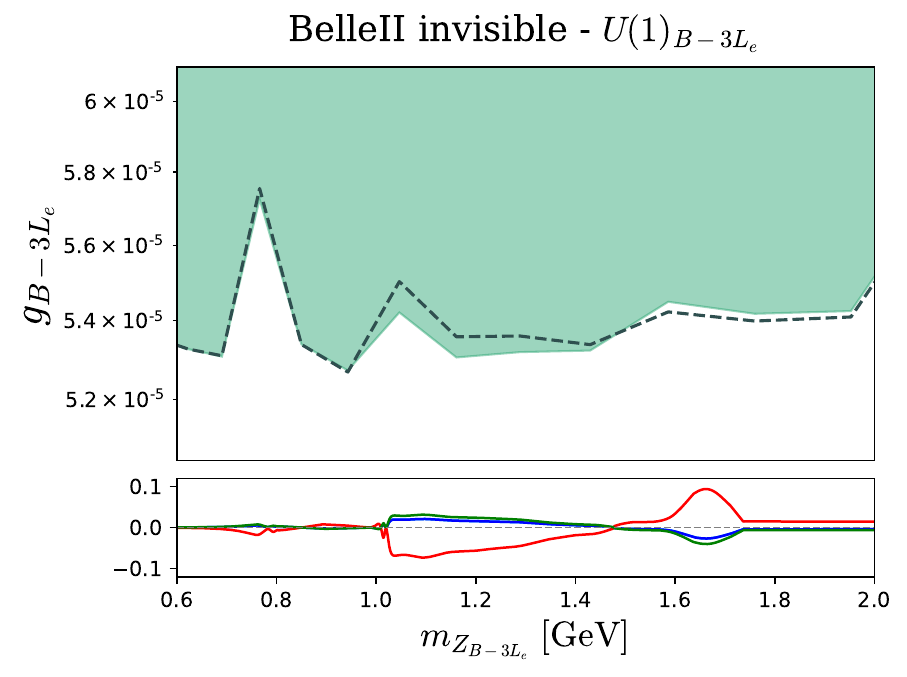}
\end{center}
\vglue -0.8cm
\caption{\label{fig:reachBelleInv}  
Expected sensitivity for the $B-L$ (left panel) and $B-3L_e$ (right panel) models for Belle II invisible searches according to our calculation (solid green) and \textsc{DarkCast} (dashed lines).
In the bottom panels we show for each model the difference of the branching ratio between the two calculations. The 11 data points shown in the figure are taken from \cite{Belle-II:2018jsg}.
 } 
\end{figure}

In figure~\ref{fig:reachBelleVis} we display the predicted sensitivity 
for Belle II (visible searches) for the $B-L$ (top left panel), $B-L_e-2L_\tau$ (top right panel) and $B-3L_e$ (bottom panel) models. We predict (solid 
light blue) for all these models a loss of sensitivity for  vector boson masses between 1.5 GeV and 1.8 GeV, due to an increase of hadronic final states (and consequent decrease of leptonic ones) in this mass window. We also predict a slight increase in sensitivity in other mass regions.

Finally, in  figure~\ref{fig:reachBelleInv} we can see the expected 
sensitivity for Belle II invisible searches for the $B-L$ (left panel) and $B-3L_e$ (right panel) models. Here again the decrease (increase) of the  hadronic final state contributions in certain mass regions, respond for the increase (decrease) of sensitivity of Belle II in the invisible mode according to our assessment.

\newpage 

\section{Final Conclusions and Outlook}
\label{sec:conc}
In this paper we  present an improved calculation of the decay width and branching ratios for baryophilic vector boson  mediators $Z_Q$, associated with a new U(1)$_Q$ gauge symmetry and having  a mass in the MeV-to-GeV range, providing for the first time an  almost  complete  set  of $Z_Q$  decays into  arbitrary  leptonic  and  hadronic final  states.

This is relevant as  one  can,  misguided  by  an  incomplete  or  incorrect theoretical description of the data, exclude regions that are still allowed and perhaps hinder the imminent discovery of a  new weak force in this mass region by future experiments. Furthermore, present and future experiments could, in principle, look for hadronic signatures of these states, in particular close to hadronic resonances.

We use a data driven approach fitting $e^+e^-$ cross-sections from 
various experiments and the meson dominance  model of chiral perturbation theory  to derive reliable predictions.
The VMD model allows us to calculate the decay widths and branching ratios of the new $Z_Q$ into light hadrons by considering its direct mixing 
to the dominant vector mesons $\rho, \omega$ and $\phi$. This was done before in \cite{Ilten:2018crw} but we improve their calculation in 
various ways.

We have updated to the most recent $e^+e^-$ data (see tables~\ref{tab:old_channels} and \ref{tab:new_channels}), we included a more complete description of the dominant vector meson contributions, we corrected the $KK$ ($\pi^0\gamma$) contribution that was overestimated (underestimated) before, we have considered the individual contributions  to the $KK\pi$ channel correctly describing the final state kinematics and  we included several new hadronic channels (see table~\ref{tab:new_channels}), in particular, above 1 GeV.
See appendix~\ref{appx:newfit} for further details on the calculations for  old and new channels.

We discussed the impact of our new calculation on the hadronic decay widths and branching ratios of some baryophilic $Z_Q$ models as well as on the current experimental limits on the plane $g_Q \times m_{{Z}_{Q}}$ for these models. 

We also show how some future experiments (FASER 2, SHiP, Belle II) can 
have their sensitivities affected by our better assessment of the $Z_Q$ hadronic modes.

The  results  for  hadronic  decays  of  any  new $Z_Q$ mediator in the MeV-to-GeV mass range are provided for public use in the python package \textsc{DeLiVeR} that is available on GitHub at \href{https://github.com/preimitz/DeLiVeR}{https://github.com/preimitz/DeLiVeR}  with a jupyter notebook  tutorial. See appendix~\ref{appx:userguide} for some information of what can be found in our hadronic decay package.

\begin{acknowledgements}
PR thanks Felix Kling for many useful discussions and for providing help with \textsc{FORESEE}. ALF and PR are supported by Funda\c{c}\~ao de Amparo \`a Pesquisa do Estado de S\~ao Paulo (FAPESP) under the contracts 2020/00174-2, and 2020/10004-7, respectively. RZF is partially supported by Funda\c{c}\~ao de Amparo \`a Pesquisa do Estado de S\~ao Paulo (FAPESP) and Conselho Nacional de Ci\^encia e Tecnologia (CNPq).
\end{acknowledgements}%

\appendix

\section{$Z_Q$: kinetic mixing, mass mixing and couplings to SM 
fermions}
\label{appx:framework}
The most general Lagrangian that describes our model includes 
the kinetic mixing between $U(1)_Y$ and $U(1)_Q$, 
direct couplings of the new $U(1)_Q$ boson  to SM fermions and mass mixing between $Z$ and $Z_Q$. The first relevant term involving the neutral gauge bosons is 
 
\be
\mathcal{L}^0_{\rm gauge} = - \frac{1}{4} \hat F_{\mu \nu} \hat F^{\mu \nu}  - \frac{1}{4} \hat Z_{Q \mu \nu} \hat Z_Q^{\mu \nu} - \frac{\epsilon}{2 \cos \theta_W} \hat Z_{Q \mu \nu} \hat F^{\mu \nu} ,
\label{eq:KM}
\ee
which describes the gauge fields $\hat{F}$ and $\hat{Z}$ mixing via the coupling of their field strength tensors $\hat{F}^{\mu \nu}$ and $\hat{Z}^{\mu \nu}_Q$. To bring this term to the canonical form 
we perform the GL(2,R) rotation
\be
\begin{pmatrix}
\hat Z_{Q}^\mu \\ 
\hat F^\mu
\end{pmatrix}
=
\begin{pmatrix}
\frac{1}{\sqrt{1-(\epsilon/c_W)^2}} & 0 \\ 
-\frac{\epsilon/c_W}{\sqrt{1-(\epsilon/c_W)^2}} & 1
\end{pmatrix}
\begin{pmatrix}
\tilde Z_{Q}^\mu \\ 
F^\mu
\end{pmatrix}\, ,
\ee
where $c_W=\cos\theta_W$, so that for $\epsilon \ll 1$ the fields are redefined as  

\be
\hat Z^\mu_Q \to \tilde Z^\mu_Q \, ,\quad \quad \hat{F}^\mu \to F^\mu - \frac{\epsilon}{c_W} \tilde Z^\mu_Q \, .
\label{eq:shift}
\ee
The interactions of the $SU(2)_L$, $U(1)_Y$ and the new  $U(1)_Q$ gauge bosons, $\hat W^{1,2,3}$, $\hat F$ and $\hat Z_Q$, respectively, with the SM chiral fermions are described by

\be
\mathcal{L}_{\rm int} = \sum_{f} i \bar{f} \gamma^\mu D_\mu f ,
\label{eq:intf}
\ee
in terms of the covariant derivative  
$$D_\mu \equiv \partial_\mu + i g 
\sum_{a=1}^{3} \frac{\tau^a}{2} \hat{W}^a_\mu + i g' Q_Y \hat{F}_\mu + i g_Q Q_{Q} \hat{Z}_{Q\mu}\, ,$$  
where $g$ ($\tau^a/2$), $g'$ ($Q_Y$) and $g_Q$ ($Q_Q$) are the $SU(2)_L$, $U(1)_Y$ and $U(1)_Q$ gauge couplings (generators).
The relevant part of this Lagrangian for the neutral sector is 
\be
\mathcal{L}^0_{\rm int} = 
- g J^\mu_3 \hat{W}^3_{\mu} - g'J^\mu_Y \hat{F}_\mu - g_Q J^{\mu}_{Q} \hat{Z}_{Q\mu}= - e J_{\rm em}^{\mu} \hat{A}_\mu - \frac{g}{c_W} J_Z^\mu \hat{Z}_\mu 
-g_Q J_{Q}^\mu \hat{Z}_{Q\mu}\, ,
\label{eq:current}
\ee
where 
\be
\begin{pmatrix}
\hat A_\mu \\ 
\hat Z_\mu
\end{pmatrix}
=
\begin{pmatrix}
s_W & c_W \\ 
c_W & -s_W
\end{pmatrix}
\begin{pmatrix}
\hat W^3_\mu \\ 
\hat F_\mu
\end{pmatrix}
=
\begin{pmatrix}
s_W & c_W \\ 
c_W & -s_W
\end{pmatrix}
\begin{pmatrix}
\hat W^3_\mu \\ 
 F_\mu - \frac{\epsilon}{c_W} \tilde Z_{Q\mu}
\end{pmatrix}
=
\begin{pmatrix}
 A_\mu - \epsilon \tilde Z_{Q\mu} \\ 
 Z_\mu + \epsilon \tan \theta_W \tilde Z_{Q\mu}
\end{pmatrix}
\, ,
\ee
and $s_W = \sin \theta_W$. We see that the transformation given by eq.(\ref{eq:shift}) introduces to $\mathcal{O}(\epsilon)$ shifts in the fields that will couple the new $\tilde Z_Q$ to the electromagnetic current $J^\mu_{\rm em}$ as well as to the neutral current $J_Z^\mu$.
However, although $A_\mu$ is already the photon field, $Z$ and $\tilde Z_Q$ are 
still not the physical fields because they are not yet the mass eigenstates. What exactly will happen depends on the structure of the extended scalar sector.
 
 To illustrate, let us consider that the scalar that breaks $U(1)_Q$ is a singlet. In this case the only source of 
 mass mixing between $\hat{Z}$ (that gets a mass $M_Z$ generated by the SM Higgs mechanism) and $\tilde Z_{Q}$ (that gets a mass $m_{\tilde Z_Q}$ generated by the singlet vacuum expectation value) is the kinetic mixing 
\bea
\mathcal{L}^0_{\rm mass}  = &
\frac{1}{2} M_Z^2 \hat{Z_\mu} \hat{Z^\mu} +\frac{1}{2} m_{\tilde Z_Q}^2 \tilde Z_{Q\mu} \tilde Z^\mu_Q\, , \nonumber \\
=& \frac{1}{2} M_Z^2 Z_\mu Z^\mu + M_Z^2 \epsilon \tan \theta_W \tilde Z_{Q\mu}Z^\mu  + 
\frac{1}{2} (m_{\tilde Z_Q}^2 + M_Z^2\epsilon^2 \tan^2\theta_W) \tilde Z_{Q\mu} \tilde Z^\mu_Q \, .
\label{eq:mass}
\eea
One can finally bring $(Z \; \tilde Z_Q)$  to 
the mass basis $(Z^{0} \;  Z_Q)$  by the following rotation 

\be
\begin{pmatrix}
Z^\mu \\ 
 \tilde Z^\mu_{Q}
\end{pmatrix}
=
\begin{pmatrix}
\cos \xi  & \sin \xi \\ 
- \sin \xi & \cos \xi
\end{pmatrix}
\begin{pmatrix}
 Z^{0\mu} \\ 
 Z^\mu_Q
\end{pmatrix}\, ,
\ee
with $$\tan 2 \xi = \frac{2 \epsilon \tan \theta_W}{1-\delta^2} + \mathcal{O}(\epsilon^2)\, , \quad {\rm with}\quad \delta^2 \equiv \frac{m^2_{\tilde Z_Q}}{M^2_Z}\, .$$
For $\epsilon, \delta \ll 1$ we have $\xi \approx - \epsilon \tan \theta_W (1+ \delta^2)$ which leads to

\be
\tilde Z^\mu_Q \to - \xi \,  Z^{0 \mu} + Z^\mu_Q \, ,\quad \quad {Z}^\mu \to Z^{0\mu} + \xi  Z^\mu_Q \, .
\label{eq:shift2}
\ee
so neglecting terms of $\mathcal{O}(\epsilon^2,\epsilon \delta, g_Q \epsilon)$

\be
\mathcal{L}^0_{\rm int} = 
 - e J_{\rm em}^{\mu} (A_\mu - \epsilon Z_{Q\mu}) - \frac{g}{c_W} J_Z^\mu Z^0_\mu 
-g_Q J_{Q}^\mu Z_{Q\mu}\, ,
\label{eq:current2}
\ee
with $M^2_{Z^0} = M^2_Z + \mathcal{O}(\epsilon^2)$ and 
$m^2_{Z_Q} = m^2_{\tilde Z_Q} + \mathcal{O}(\epsilon^2)$.

If the scalar that breaks $U(1)_Q$ is an $SU(2)_L$ doublet or triplet
there are going to be other sources of mass mixing.

\section{Details of the hadronic fit calculation}
\label{appx:newfit}

In this appendix we provide additional details concerning the hadronic calculation described in section \ref{sec:HadCalc}.

\paragraph{Dominant Low-Energy Hadronic Modes:}
In section \ref{sec:HadCalc}, we highlight  the improvements obtained by using our VMD calculation compared to the old channels already included in \textsc{DarkCast}. Specially for the case of the $\pi^0 \gamma$, $KK$ and $KK\pi$ channels we show that considerable differences appear. These divergences arise mainly because of the inclusion of other vector meson components, but also as a result of the use of additional data in our fits. 

\begin{figure}[htb]
\begin{center}
\includegraphics[width=0.49\textwidth]{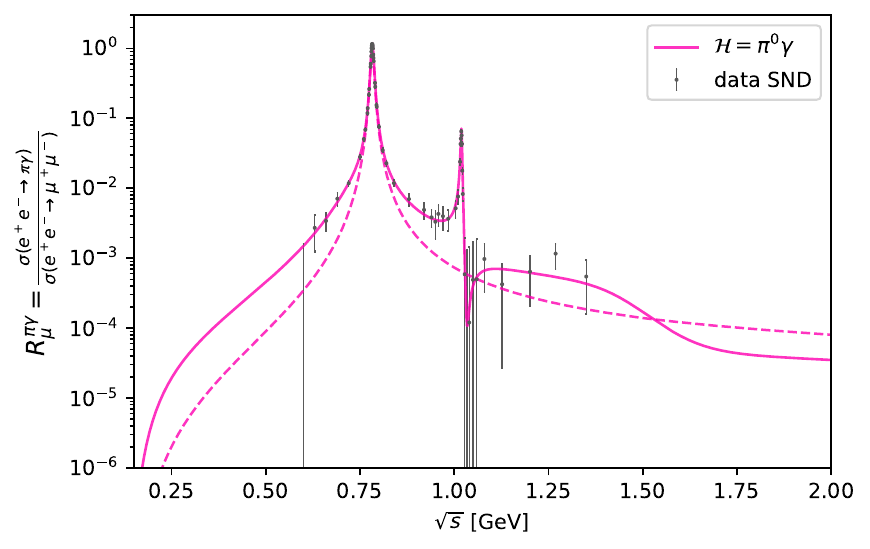}
\includegraphics[width=0.49\textwidth]{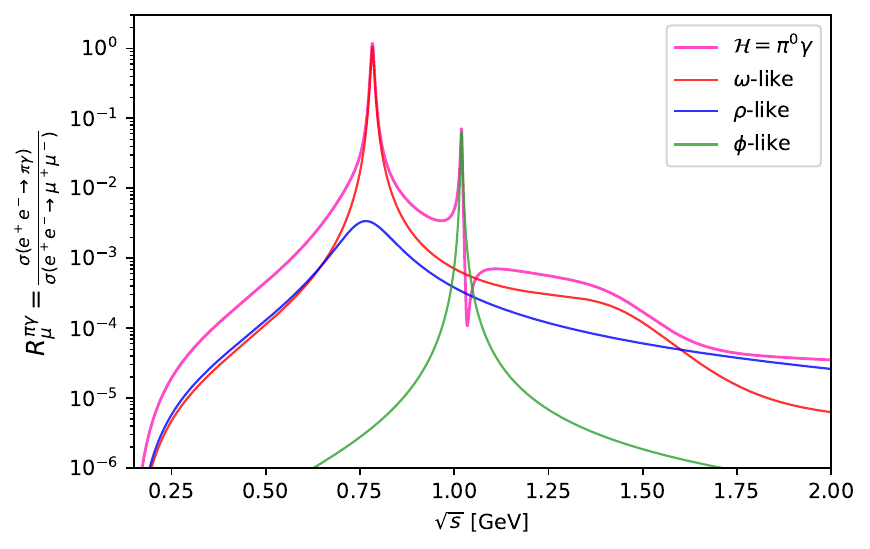}
\end{center}
\vglue -0.8cm
\caption{\label{fig:Rpigamma} Normalized cross-section $R^{\mathcal{H}}_\mu$ for the 
$\mathcal{H}=\pi^0 \gamma$ channel. In the left panel the solid (dashed) pink line indicates our (\textsc{DarkCast}) calculation, and the gray data points are from SND \cite{SND:2016drm}. In the right panel we show our decomposition of the $\pi \gamma$ channel into $\rho$ (blue), $\omega$ (red) and $\phi$ (green) components.
  }
\end{figure}

In the left panel of figure \ref{fig:Rpigamma}, we show our (\textsc{DarkCast}) $R$-ratio calculation for the $\pi \gamma$ channel in solid (dashed) pink together with the experimental data points from the SND collaboration \cite{SND:2016drm}. In the right panel of the same figure, we show the decomposition of the $\pi \gamma$ ratio into $\rho$ (blue), $\omega$ (red) and $\phi$ (green) contributions. From the figure, we can see that the second peak close to $1$ GeV comes from a $\phi$-like component that was not included in \textsc{DarkCast}. The dip that appears right after this peak is a consequence of the interference term between $\omega$ and $\phi$ contributions. Altogether, we can see that the inclusion of all the vector meson components provides a better description of the data points.

\begin{figure}[htb]
\begin{center}
\includegraphics[width=0.49\textwidth]{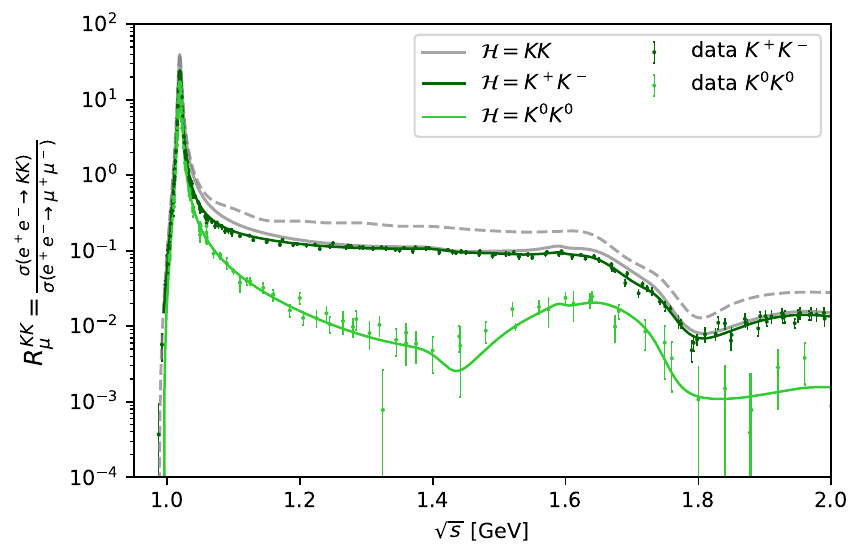}
\includegraphics[width=0.49\textwidth]{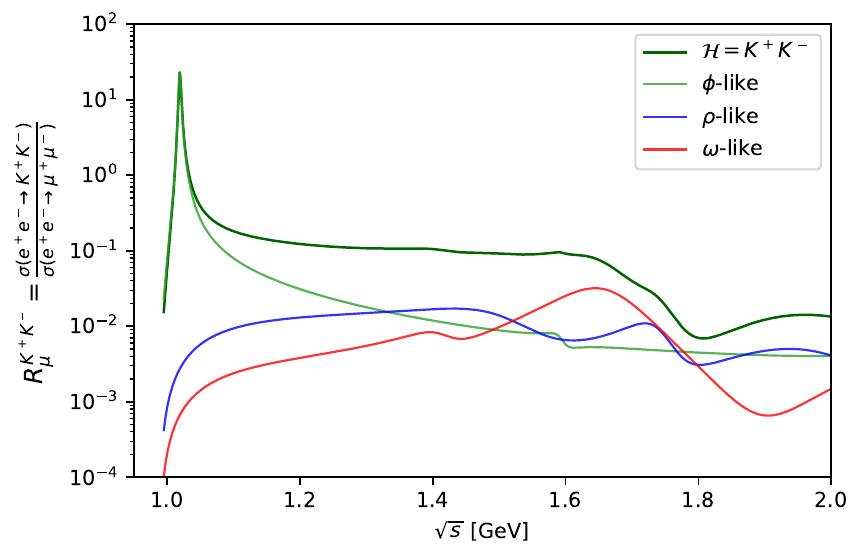}
\end{center}
\vglue -0.8cm
\caption{\label{fig:RKK} Normalized cross-section $R^{\mathcal{H}}_\mu$ for the $\mathcal{H}=KK$ channel. In the left panel the solid (dashed) gray line indicates our (\textsc{DarkCast}) total $KK$ calculation, while the dark (light) green line indicated our fit for the $K^+K^-$ ($K^0 \bar{K}^0$) channel. The data points correspond to a compilation from several experiments \cite{CLEO:2005tiu,Achasov:2000am,Achasov:2006bv,Mane:1980ep,CMD-3:2016nhy,BaBar:2014uwz,CMD-2:2008fsu,BaBar:2013jqz,BaBar:2015lgl,Achasov:2016lbc}. In the right panel we show the decomposition of the charged $K^+K^-$ channel (dark green) into $\rho$ (blue), $\omega$ (red) and $\phi$ (green) components.}
\end{figure}

For the case of the $KK$ channel, in the left panel of figure \ref{fig:RKK} we show the individual normalized cross sections for the neutral $K^0\bar{K}^0$ (light green) and charged $K^+K^-$(dark green) channels obtained using the fit from \cite{Plehn:2019jeo}, along with the corresponding data points extracted from \cite{CLEO:2005tiu,Achasov:2000am,Achasov:2006bv,Mane:1980ep,CMD-3:2016nhy,BaBar:2014uwz,CMD-2:2008fsu,BaBar:2013jqz,BaBar:2015lgl,Achasov:2016lbc}. In solid (dashed) grey we show our (\textsc{DarkCast}) total $KK$ contribution, where for our calculation $KK = K^0 \bar{K}^0 + K^+K^-$, while in \textsc{DarkCast} $KK=2\; K^+K^-$. In the latter case, the reason for this definition is a consequence of the exclusive use of BaBar data from \cite{BaBar:2013jqz}, which was a study that considered only the charged channel contribution. Here, we update the $KK$ channel description also by including recent data from several experiments.

In the right panel of figure \ref{fig:RKK}, we present the decomposition of the charged $K^+K^-$ channel (dark green) into $\rho$ (blue), $\omega$ (red) and $\phi$ (green) components. It is important to emphasize that in \textsc{DarkCast} only the $\phi$-like component, that is responsible for the peak near $1$ GeV, is considered. However, the other features of the fit mainly come from the remaining vector meson contributions included in this study.

\begin{figure}[htb]
\begin{center}
\includegraphics[width=0.31\textwidth]{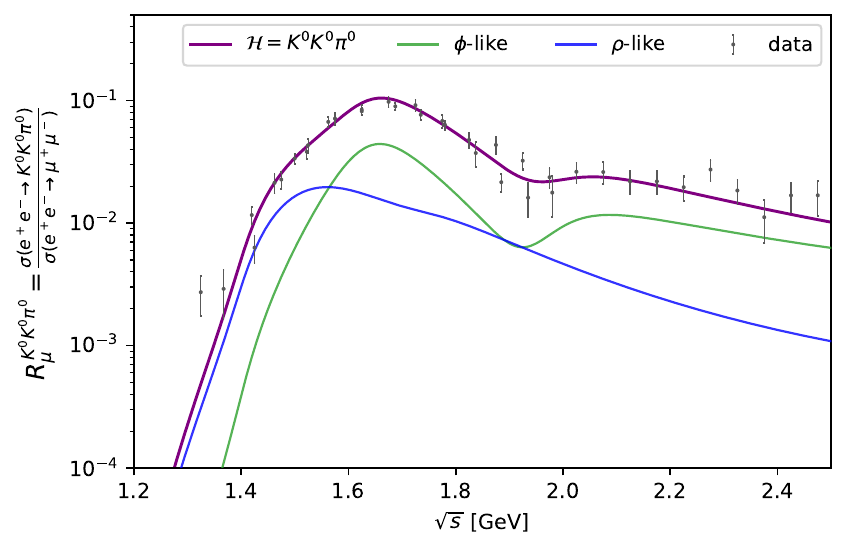}
\includegraphics[width=0.31\textwidth]{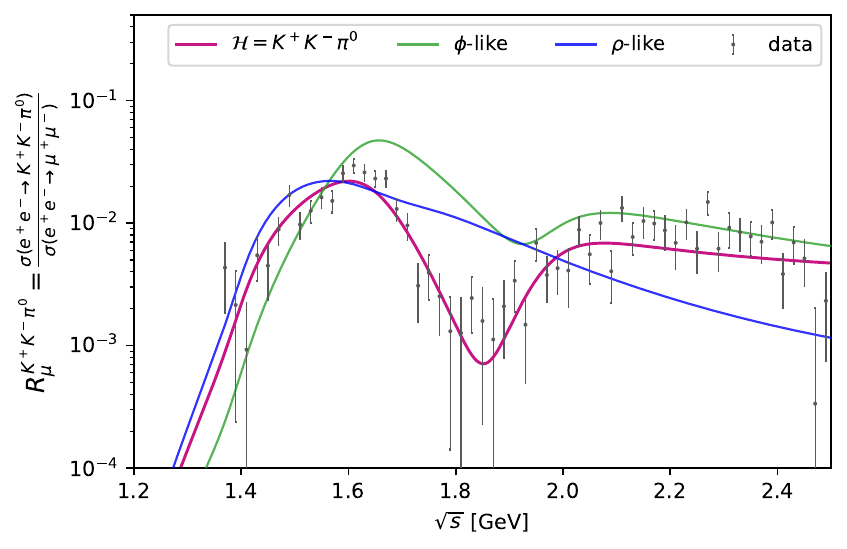}
\includegraphics[width=0.31\textwidth]{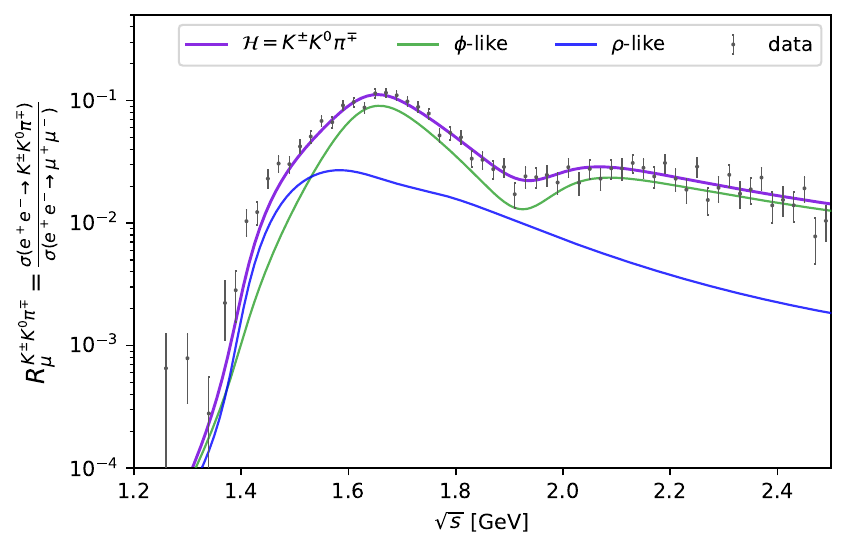}
\end{center}
\vglue -0.8cm
\caption{\label{fig:RKKpi} Normalized cross-section $R^{\mathcal{H}}_\mu$ for $\mathcal{H}=K^0K^0\pi^0$(left), $K^+K^-\pi^0$ (middle), $K^\pm K^0\pi^\mp$ (right). The purple lines correspond to the channel contribution, whereas the blue and green lines indicate the $\phi$ and $\rho$ decomposition, respectively. The data points from \cite{BaBar:2007ceh,BaBar:2017nrz,Achasov:2017vaq,Bisello:1991kd,Mane:1982si} are shown in grey.}
\end{figure}

Finally, figure \ref{fig:RKKpi} shows the normalized cross section obtained using the fit from \cite{Plehn:2019jeo} for the individually $KK\pi$ channels, together with the decomposition into $\phi$ (green) and $\rho$ (blue) contributions and the data points extracted from \cite{BaBar:2007ceh,BaBar:2017nrz,Achasov:2017vaq,Bisello:1991kd,Mane:1982si}. The sum of these three channels results in the total $KK\pi$ channel considered in this work, in contrast to the $KK \pi$ used by \textsc{DarkCast}, which consists only of the isoscalar component and agrees with a different set of data \cite{BaBar:2007ceh}. Hence, we not only consider a new vector component to the $KK\pi$ channel, but also used more recent data and described it correctly including separately the three components $K^0K^0\pi^0$, $K^+K^-\pi^0$ and $K^\pm K^0\pi^\mp$.

\paragraph{New channels:} Besides the additional channels described in \cite{Plehn:2019jeo} that we include in this study, we add the description of four new channels, relevant in the higher energy region close to $2$ GeV. For the inclusion of these channels, first we need to identify all the possible intermediate structures. Then, if the data points are available, we can perform a fit using the python package \verb!IMinuit! \cite{hans_dembinski_2021_5561211}. Below, we provide additional details concerning the method used for the computation of the fit for each of these new channels.
\begin{itemize}
    \item \textbf{$\mathcal{H} = \omega \pi \pi$}

In case of the $\omega\pi\pi$ final state, we distinguish between the
charged mode $\omega\pi^+\pi^-$ and the neutral mode $\omega\pi^0\pi^0$, dominantly leading to the five pion final state combinations $2(\pi^+\pi^-)\pi^0$ and $\pi^+\pi^-3\pi^0$, respectively. Although we have signs of possible intermediate substructures, such as $\omega f_0(980)\to \omega \pi^+\pi^-$ \cite{Aubert:2007ef} and $b_1(1285)\pi\to \omega\pi\pi$ \cite{Achasov:1998sv}, so far they have not been clearly seen in the data and hence, we will not consider them. 

Considering that G-parity only allows for $I=0$, we assume a point-like $\omega \to \omega \pi\pi$ interaction. The form factor is given by
\begin{align}
F_{\omega\pi\pi} = \sum_V \frac{a_V m_V^2 e^{i\varphi_V}}{m_V^2-s-i\sqrt{s}\Gamma_V} \, ,
\end{align}
where the only vector meson relevant to describe this channel is $V=\omega''$, which corresponds to the $\omega(1650)$ meson. For the fit, we use data from \cite{Akhmetshin:2000wv,Aubert:2007ef,Lees:2018dnv}. Table \ref{tab:omegapipi} lists the fit parameters obtained for this channel and figure \ref{fig:fitomegapipi} shows the curve of the fit with the hadronic data.

\begin{table}
\def\arraystretch{1.2}
  \begin{center}
    \begin{tabular}{|c|c||c|c|}
      \hline
      Parameter & Fit Value & Parameter & Fit Value \\
      \hline
      $m_{\omega''}$ & $1.661\pm0.007$\,GeV & $\Gamma_{\omega''}$ & $0.398\pm0.021$\,GeV \\
      $a_{\omega''}$ & $2.73\pm0.09$ & $\varphi_{\omega''}$ & $0$ (fixed) \\
    \hline 
    \multicolumn{4}{|c|}{$\chi^2 \mathrm{/n.d.f.} = 1.67$} \\
    \hline
    \end{tabular}
    \end{center}
    \caption{Values obtained by the fit for the $e^+e^-\to \omega \pi \pi$ current. The phase $\varphi_{\omega''}$ was fixed at $0$.}
    \label{tab:omegapipi}
  \end{table}
  
\begin{figure}[htb]
\begin{center}
\includegraphics[width=0.45\textwidth]{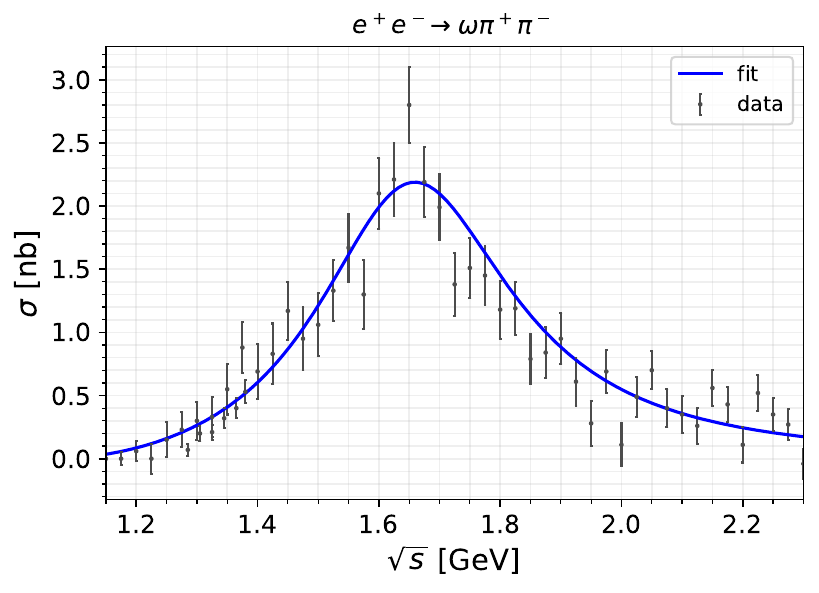}
\includegraphics[width=0.45\textwidth]{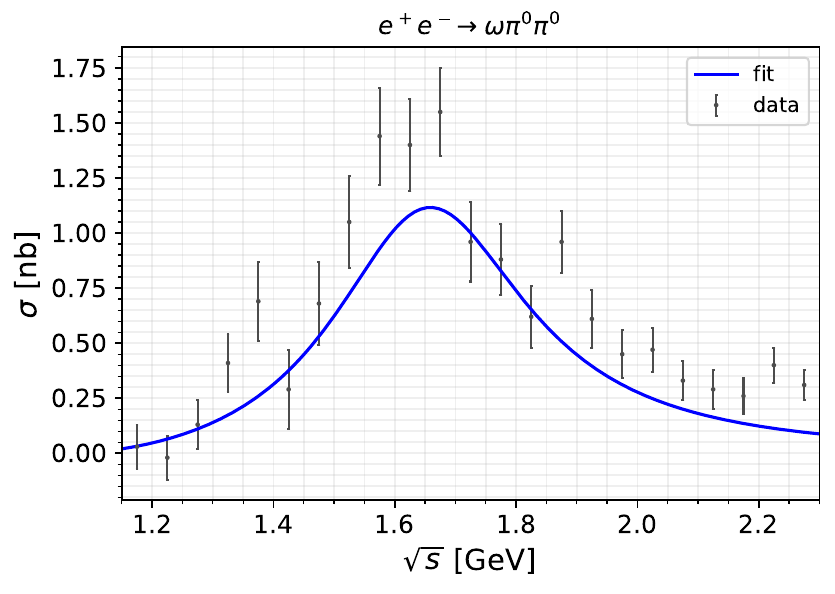}
\end{center}
\vglue -0.8cm
\caption{\label{fig:fitomegapipi} Cross-section for the charged $\omega\pi^+\pi^-$ (left panel) and neutral $\omega\pi^0\pi^0$ (right panel) hadronic final states. The blue curve shows the best fit solution to the cross-section, obtained considering the fit values of table \ref{tab:omegapipi}. The black points and error bars represent data from \cite{Akhmetshin:2000wv,Aubert:2007ef,Lees:2018dnv}.  }
\end{figure}

    \item \textbf{$\mathcal{H} = K^*(892) K \pi$}

The $K^*(892) K \pi$ channel is the dominant contribution to $KK\pi\pi$ final states. Here, we consider the four most relevant intermediate substructures: $K^{*0} K^\pm \pi^\mp$, $K^{* \pm} K_S \pi^\mp$ and $K^{* \pm} K^\mp \pi^0$ decaying into $K_S K^\pm \pi^\mp \pi^0$, and $K^{*0} K^\pm \pi^\mp$ decaying into $K^+ K^- \pi^+ \pi^-$. 

In order to calculate the form factors we need to combine the isospin $I=0$ ($\phi$) and $I=1$ ($\rho$) contributions, corresponding to $A_0$ and $A_1$ amplitudes, respectively. The most general form factors assuming a point-like $V\to K^{*0}K^\mp \pi^\pm$ vertex structure can be written as
\begin{align}
K^{*0} K^\pm \pi^\mp \to K_S K^\pm \pi^\mp \pi^0: F_{(K^{*0} K^\pm \pi^\mp)^n} = \frac{1}{\sqrt{18}}(A_1+A_0)\notag\\
K^{*\pm} K_S \pi^\mp \to K_S K^\pm \pi^\mp \pi^0: F_{K^{* \pm} K_S \pi^\mp} = \frac{1}{\sqrt{18}}(A_1-A_0)\notag\\
K^{* \mp} K^\pm \pi^0 \to K_S K^\pm \pi^\mp \pi^0: F_{K^{* \pm} K^\mp \pi^0} = \frac{1}{\sqrt{18}}(A_1-A_0)\notag\\
K^{*0} K^\pm \pi^\mp \to K^+ K^- \pi^+ \pi^-: F_{(K^{*0} K^\pm \pi^\mp)^c} = \sqrt{\frac{2}{9}}(A_1+A_0)~,\notag
\label{eq:KKpipi}
\end{align} 
where we denote the form factor of the $K^{*0} K^\pm \pi^\mp$ state decaying into the neutral (charged) $K_S K^\pm \pi^\mp \pi^0$ ($K^+ K^- \pi^+ \pi^-$) with a \textit{n} (\textit{c}) superscript and the isospin amplitudes are given by
\begin{align}
A_0&=\sum_\phi
     \frac{a_{\phi}e^{i\varphi_\phi}m_\phi^2}{m_\phi^2-s-i
     m_\phi\Gamma_\phi}~,\notag\\
A_1&=\sum_\rho
     \frac{a_{\rho}e^{i\varphi_\rho}m_\rho^2}{m_\rho^2-s-i m_\rho\Gamma_\rho}~.
\end{align}

The data used to perform the fit was taken from BaBar \cite{BaBar:2011btv,BaBar:2017pkz}, and the vector meson resonances found by the fit to describe this channel were $V= \phi', \rho''$. Table \ref{tab:KKpipi} summarizes the obtained fit parameters and figure \ref{fig:fitKKpipi} shows the curve of the best fit solution for each of these four $KK\pi\pi$ states.

\begin{table}
\def\arraystretch{1.3}
  \begin{center}
    \begin{tabular}{|c|c||c|c|}
      \hline
      \multicolumn{2}{|c||}{$K_S K^\pm \pi^\mp \pi^0$} & \multicolumn{2}{|c|}{$K^+ K^- \pi^+ \pi^-$} \\
      \hline
      \hspace{1pt} Parameter \hspace{1pt} & \hspace{20pt} Fit Value \hspace{20pt} & \hspace{1pt} Parameter \hspace{1pt} & {\hspace{20pt} Fit Value \hspace{20pt}} \\
      \hline
      $m_{\phi'}$ & $1.7$\,GeV (fixed) & $m_{\phi'}$ & $1.65$\,GeV  (fixed)\\
      $a_{\phi'}$ & $2.49\pm0.6$ & $a_{\phi'}$ & $4.52\pm0.5$ \\
      $\Gamma_{\phi'}$ & $0.3$\,GeV (fixed) & $\Gamma_{\phi'}$ & $0.103\pm0.009$\\
       $\varphi_{\phi'}$ & $1.02\pm0.09$ & $\varphi_{\phi'}$ & $\pi$ (fixed) \\
      $m_{\rho''}$ & $1.898\pm0.012$\,GeV  & $m_{\rho''}$ & $ 1.842 \pm 0.011$\,GeV  \\
      $a_{\rho''}$ & $13.5 \pm 0.5$ & $a_{\rho''}$ & $15.7 \pm 1.0$ \\
      $\Gamma_{\rho''}$ &  $0.504 \pm  0.021$\,GeV  & $\Gamma_{\rho''}$ &  $0.403 \pm  0.016$\,GeV   \\
       $\varphi_{\rho''}$ & $0$ (fixed) & $\varphi_{\rho''}$ & $0.000 \pm 0.006$ \\       
    \hline 
          \multicolumn{2}{|c||}{$\chi^2 \mathrm{/n.d.f.} = 2.68$} & \multicolumn{2}{|c|}{$\chi^2 \mathrm{/n.d.f.} = 1.59$} \\
    \hline
    \end{tabular}
    \end{center}
    \caption{Values obtained by the fit to the $e^+e^- \to K^* K \pi \to K_S K^\pm \pi^\mp \pi^0$ current (left) and to the $e^+e^- \to K^* K \pi \to K^+ K^- \pi^+ \pi^-$ current (right).}
    \label{tab:KKpipi}
  \end{table}

\begin{figure}[htb]
\begin{center}
\includegraphics[width=0.41\textwidth]{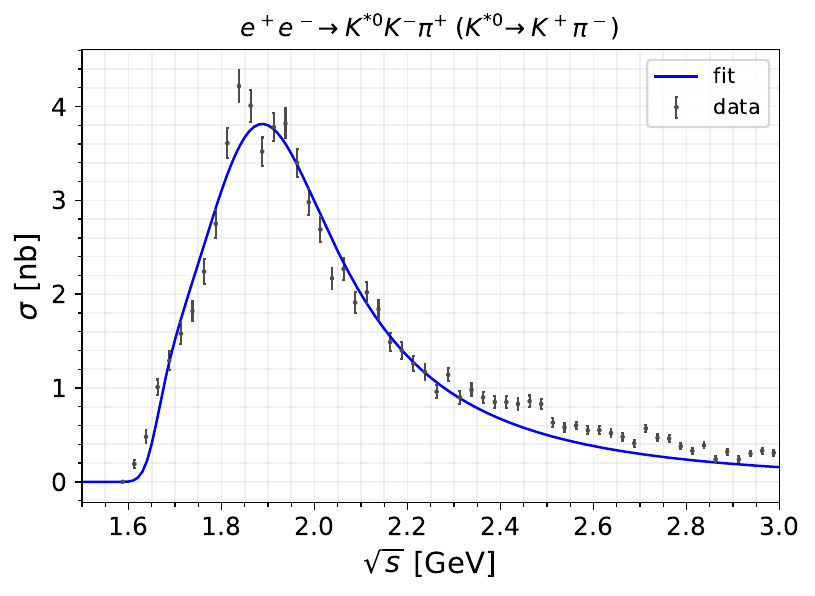}
\includegraphics[width=0.42\textwidth]{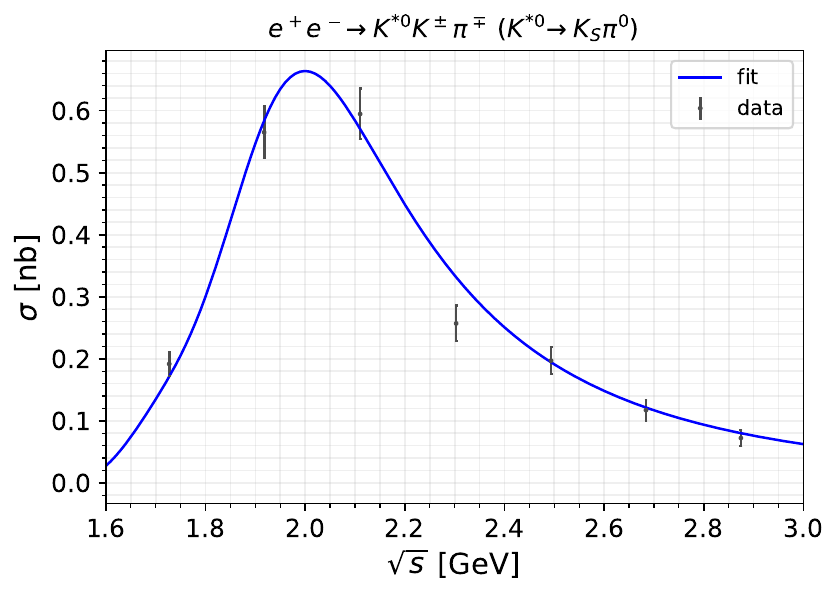}
\includegraphics[width=0.42\textwidth]{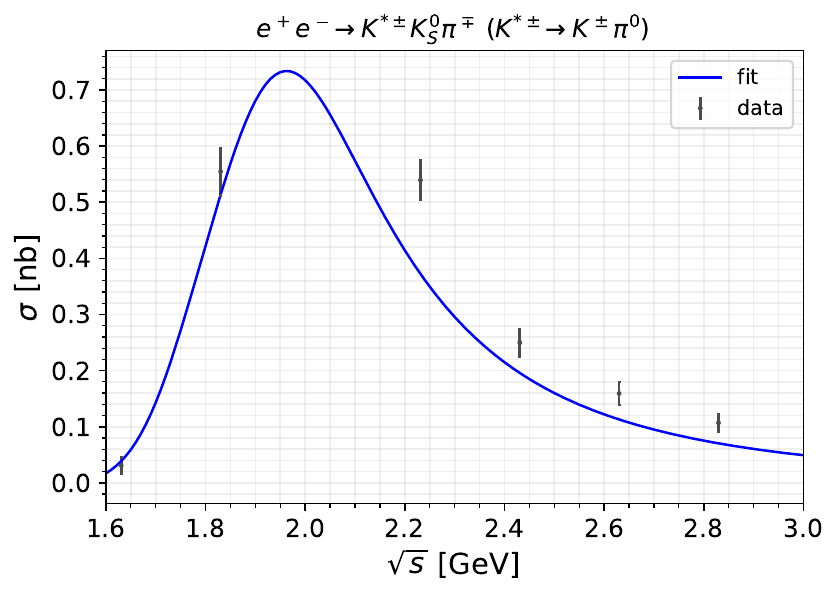}
\includegraphics[width=0.42\textwidth]{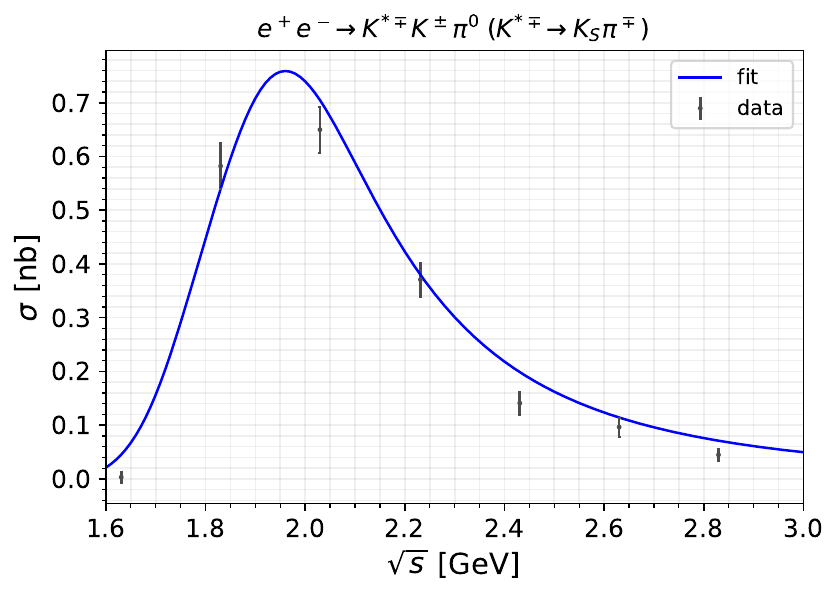}
\end{center}
\vglue -0.8cm
\caption{\label{fig:fitKKpipi} Cross-section for the charged $K^{*0} K^\pm \pi^\mp$ (upper left panel), neutral $K^{*0} K^\pm \pi^\mp$ (upper right panel), $K^{*\pm} K_S \pi^\mp$ (lower left panel) and $K^{* \mp} K^\pm \pi^0$ (lower right panel) hadronic final states. The blue curve shows the best fit solution to the cross-section, obtained considering the fit values of table \ref{tab:KKpipi}. The black points and error bars represent data from \cite{BaBar:2011btv,BaBar:2017pkz}.}
\end{figure}

    \vspace{0.8cm}
    \item \textbf{$\mathcal{H} = \phi \pi \pi$}

The $\phi \pi \pi$ channel can be decomposed into a charged ($\phi \pi^+ \pi^-$) and a neutral ($\phi \pi^0 \pi^0$) component. Due to the decay of the $\phi$ meson into two kaons, $\phi(1020) \to K^+ K^-$, the charged (neutral) component represents a small contribution to the   $K^+K^-\pi^+\pi^-$ ($K^+K^-\pi^0\pi^0$) state.

In the energy region relevant in this study, we can describe both $\phi \pi \pi$ states by the first two excited $\phi$ resonances $\phi'$ and $\phi''$. The form factor is given by
\begin{align}
F_{\phi\pi\pi} = \sum_V \frac{a_V m_V^2 e^{i\varphi_V}}{m_V^2-s-i\sqrt{s}\Gamma_V}~,
\end{align}
with $V=\phi',\phi''$. The data considered for the fit was taken from  \cite{BaBar:2011btv,Belle:2008kuo} and the values for the parameters obtained in the fit, for both the neutral and charged states, can be found in table \ref{tab:phipipi}. Figure \ref{fig:fitphipipi} shows the curve of the best fit together with the data points. 

\begin{table}
\def\arraystretch{1.2}
  \begin{center}
    \begin{tabular}{|c|c||c|c|}
      \hline
      \hspace{1pt} Parameter \hspace{1pt} & \hspace{20pt} Fit Value \hspace{20pt} & \hspace{1pt} Parameter \hspace{1pt} & {\hspace{20pt} Fit Value \hspace{20pt}} \\
      \hline
      $m_{\phi'}$ & $1.680\pm 0.012$\,GeV & $m_{\phi''}$ & $2.162\pm0.015$\,GeV \\
      $a_{\phi'}$ & $1.44\pm0.16$ & $a_{\phi''}$ & $0.69 \pm 0.11$ \\
    $\Gamma_{\phi'}$ & $0.226\pm0.021$\,GeV & $\Gamma_{\phi''}$ & $0.209\pm0.025$\,GeV \\
      $\varphi_{\phi'}$ & $2.6\pm0.4$ & $\varphi_{\phi''}$ & $0$ (fixed) \\
    \hline 
    \multicolumn{4}{|c|}{$\chi^2 \mathrm{/n.d.f.} = 0.55$} \\
    \hline
    \end{tabular}
    \end{center}
    \caption{Values obtained by the fit for the $e^+e^-\to \phi \pi \pi$ current.}
    \label{tab:phipipi}
  \end{table}

\begin{figure}[htb]
\begin{center}
\includegraphics[width=0.43\textwidth]{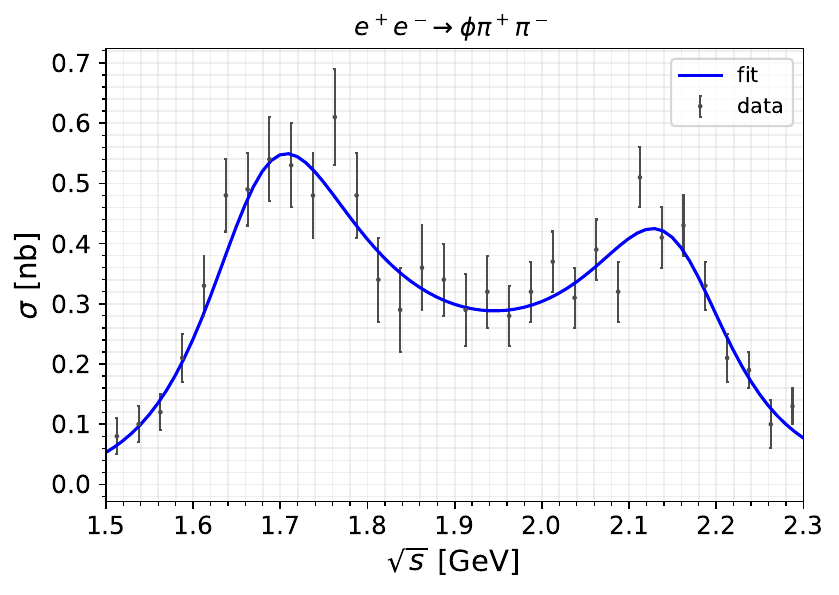}
\includegraphics[width=0.45\textwidth]{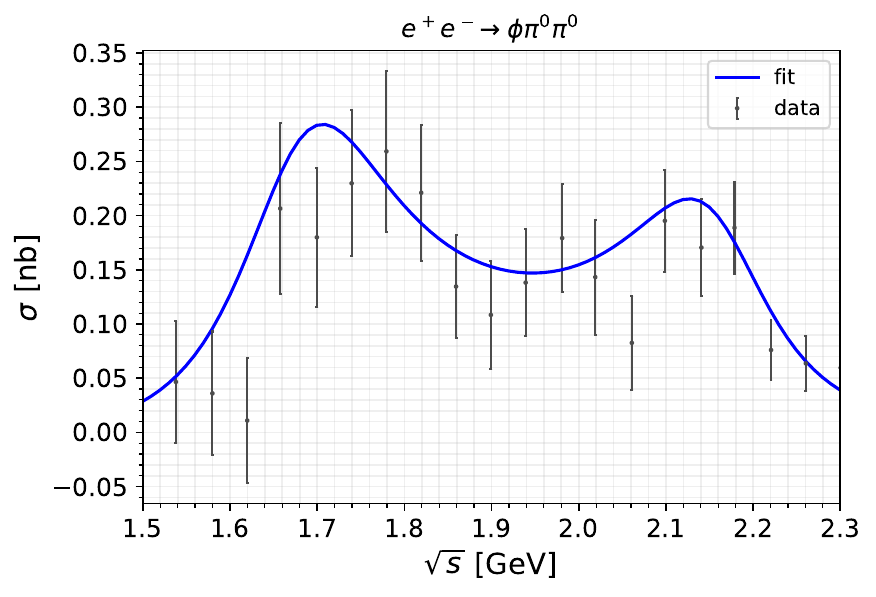}
\end{center}
\vglue -0.8cm
\caption{\label{fig:fitphipipi} Cross-section for the charged $\phi \pi^+ \pi^-$ (left panel) and neutral $\phi\pi^0\pi^0$ (right panel) hadronic final states. The blue curve shows the best fit solution to the cross-section, obtained considering the fit values of table \ref{tab:phipipi}. The black points and error bars represent data from  \cite{BaBar:2011btv,Belle:2008kuo}.  }
\end{figure}

    \item \textbf{$\mathcal{H} = 6 \pi$}
    
    For the case of the $6 \pi$ channel, we considered a charged state $3(\pi^+ \pi^-)$ and a neutral state $2(\pi^+\pi^- \pi^0)$. In this particular case, we cannot describe these channels by identifying the Breit-Wigner resonances. The available data does not indicate any clear intermediate structures, suggesting that the description of the $6\pi$ channel can only proceed via the inclusion of the decays of many different vector states. Following \cite{BaBar:2006vzy,Achasov:1996gw}, we fitted the $6\pi$ cross-section according to the parametrization given by
    \be
    \label{eq:par6pia}
    \sigma_{6\pi} = \frac{4\pi \alpha^2}{s^{3/2}} \left( \frac{a_V m_V^2 e^{i\varphi_V}}{s-m_V^2 +i \sqrt{s} \Gamma_V} + A_{\rm cont} \right)^2 \, ,
    \ee 
    where 
    \be
    \label{eq:par6pib}
    A_{\rm cont} = c_0 + c_1 \frac{e^{-b/(\sqrt{s}-m_0)}}{(\sqrt{s}-m_0)^{2-a}}
    \ee
    is a Jacob-Slansky amplitude \cite{PhysRevD.5.1847} that accounts for the mixture of several broad resonances, and the parameters $c_0,c_1,a,b,m_0$ are free variables. Due to G-parity symmetry arguments we can identify $V$ with the higher excitation of the $\rho$ vector meson, namely $V= \rho'''$. Table \ref{tab:6pi} shows the values of the free parameters obtained by our fit using the parameterization described above and the data from \cite{BaBar:2006vzy}. Figure \ref{fig:fit6pi} shows the best fit solution to the cross-section for both the charged and neutral $6\pi$ states.
    
    \begin{table}
    \def\arraystretch{1.3}
     \begin{center}
     \begin{tabular}{|c|c||c|c|}
      \hline
      \multicolumn{2}{|c||}{$3(\pi^+ \pi^-)$} & \multicolumn{2}{|c|}{$2(\pi^+\pi^- \pi^0)$} \\
      \hline
      \hspace{1pt} Parameter \hspace{1pt} & \hspace{15mm} Fit Value \hspace{15mm} & \hspace{1pt} Parameter \hspace{1pt} & {\hspace{15mm} Fit Value \hspace{15mm}} \\
      \hline
      $m_{\rho'''}$ & $1.88$\,GeV (fixed) & $m_{\rho'''}$ & $1.86$\,GeV  (fixed)\\
      $a_{\rho'''}$ & $0.0037 \pm 0.0005 \, \mathrm{GeV}^{1/2}$ & $a_{\rho'''}$ & $-0.0072 \pm 0.0009 \, \mathrm{GeV}^{1/2}$ \\
      $\Gamma_{\rho'''}$ & $0.13$\,GeV (fixed) & $\Gamma_{\rho'''}$ &  $0.16$\,GeV (fixed)\\
       $\varphi_{\rho'''}$ & $0.367$ (fixed) & $\varphi_{\rho'''}$ & $-0.052$ (fixed) \\
      $c_0$ & $0.0153\pm0.0029\, \mathrm{GeV}^{1/2}$   & $c_0$ & $ -0.028 \pm 0.009\, \mathrm{GeV}^{1/2}$   \\
      $c_1$ & $ -1.082 \pm 0.017 \, \mathrm{GeV}^{(5/2 -a)}$  & $c_1$ & $2.4 \pm 0.7\, \mathrm{GeV}^{(5/2 -a)}$ \\
      $b$ &  $1.40 \pm  0.01$\,GeV  & $b$ &  $1.54 \pm  0.32$\,GeV   \\
      $a$ & $ 0.89 \pm 0.04$ & $a$ & $0.85 \pm 0.23$ \\   
      $m_0$ & $1.262 \pm 0.012$\,GeV  & $m_0$ & $1.20 \pm 0.04$\,GeV \\     
    \hline 
          \multicolumn{2}{|c||}{$\chi^2 \mathrm{/n.d.f.} = 0.6$} & \multicolumn{2}{|c|}{$\chi^2 \mathrm{/n.d.f.} = 0.7$} \\
    \hline
    \end{tabular}
    \end{center}
    \caption{Fit values for the $e^+e^- \to 3(\pi^+ \pi^-)$ current (left) and for the $e^+e^- \to 2(\pi^+\pi^-)\pi^0\pi^0$ current (right). The values of 
    $m_{\rho'''}$, $\Gamma_{\rho'''}$ and $\varphi_{\rho'''}$ were taken from \cite{BaBar:2006vzy}.}
    \label{tab:6pi}
  \end{table}
 
\begin{figure}[htb]
\begin{center}
\includegraphics[width=0.45\textwidth]{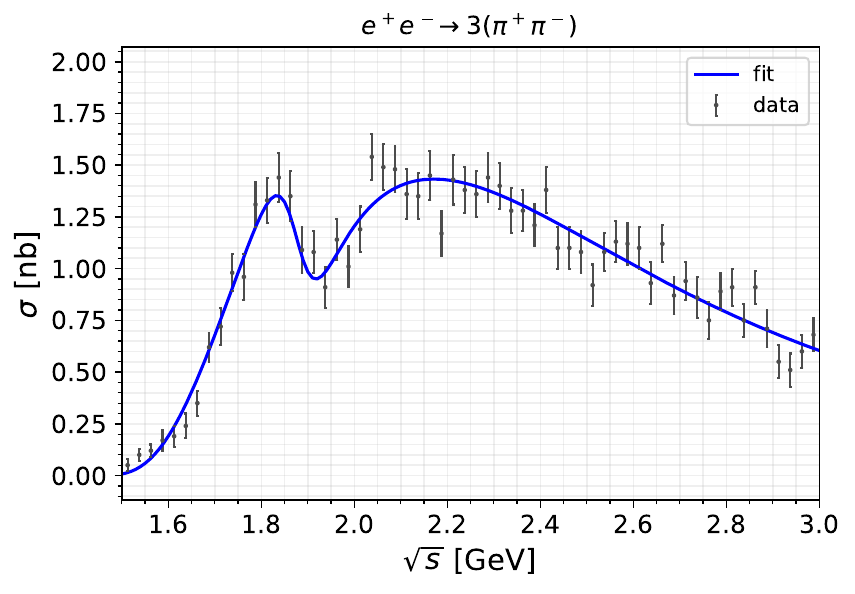}
\includegraphics[width=0.435\textwidth]{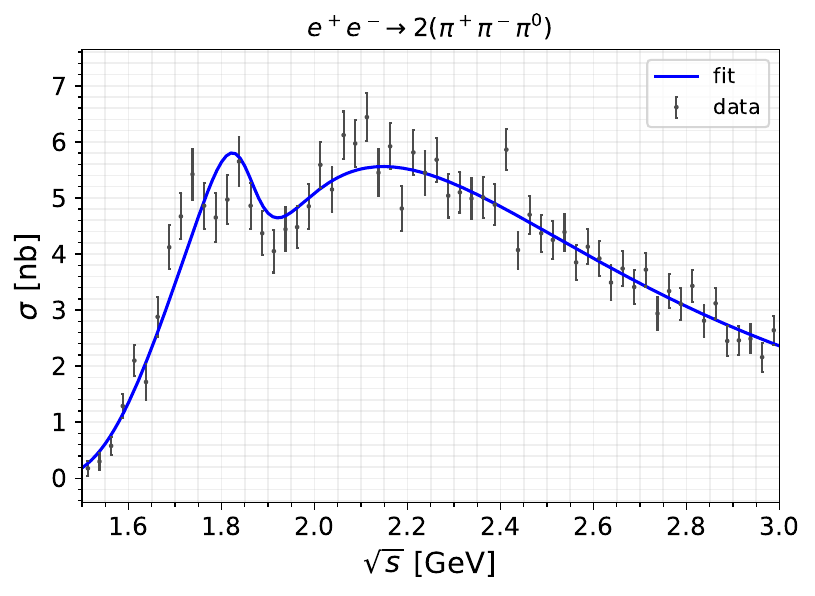}
\end{center}
\vglue -0.8cm
\caption{\label{fig:fit6pi}  Cross-section for the charged $3(\pi^+ \pi^-)$ (left panel) and neutral $2(\pi^+\pi^- \pi^0)$ (right panel) hadronic final states. The blue curve shows the best fit solution to the cross-section, obtained considering the parameterization described in eq.~(\ref{eq:par6pia}) and (\ref{eq:par6pib}) and the fit values of table \ref{tab:6pi}. The black points and error bars represent data from \cite{BaBar:2006vzy}.}
\end{figure}

\end{itemize}

\section{Hadronic Decay Package}
\label{appx:userguide}
We provide the results of calculating hadronic decays and related quantities in the python package \href{https://github.com/preimitz/DeLiVeR}{DeLiVeR}. Here, we briefly describe the structure of the code. Further instructions and more practical advices are provided in the package as jupyter notebooks.

\paragraph{Model definition:} As a first step, the user must define the model by specifying \textit{i)} the $Q$-charges $q_Q^f$ and the $U(1)_Q$ coupling constant $g_Q$, \textit{ii)} if the mediator is coupling to DM, and if yes, in which way. Possible candidates are a Majorana~\cite{Batell:2021blf,Batell:2021aja}, or Dirac fermion~\cite{MiniBooNEDM:2018cxm}, a complex scalar particle~\cite{MiniBooNE:2017nqe,MiniBooNEDM:2018cxm,DeRomeri:2019kic,Batell:2021blf,Batell:2021aja}, or inelastic pseudo-Dirac DM~\cite{Berlin:2018bsc,Batell:2021ooj}. For all DM types, the DM mass $m_\chi$ is set in relation to the mediator mass via $R_\chi=m_\chi/m_{Z_Q}$ as it is common practice in the literature and the mediator-DM coupling strength can be specified as well. In case of inelastic DM, the mass splitting between the DM states has to be defined as well. As we focus on hadronic decays of vector particles, we refrain from including an invisible DM decay of vector mediators in the results presented above and leave this study for future work.

\paragraph{Width calculation:} In order to calculate the decay widths of the $Z_Q$ mediator in the specified model, the width class is initiated by using the model as an input. As described in section~\ref{sec:HadCalc}, some channels $\mathcal{H}$ are sums of several final state configurations. For example $\mathcal{H}=4\pi$ consists of $2\pi^+2\pi^-$ and $\pi^+\pi^-2\pi^0$. Additionally to the summed contributions, the user can calculate single sub-contributions within the width class. Besides all contributions, the total width and lifetime of the particle is calculated without further ado.

\paragraph{Branching ratios:} All calculated widths can then be used to calculate the branching ratios for all channels, as well as to determine the visible, invisible and hadronic width of the vector mediator.

\paragraph{R-ratio:} In the style of the typical way of calculating the dark photon decay width as in eq.~(\ref{eq:widthDP}), the user can calculate the $R_\mu^\mathcal{H}$ values for arbitrary vector mediator models. Multiplied by the decay width $\Gamma_{Z_Q \to \mu^+\mu^-}$ this yields the partial decay width $\Gamma_{Z_Q \to \mathcal{H}}$. This method is most useful for the \textit{dark photon}, but can in principle also be used by other models. In this case the $R_\mu^\mathcal{H}$ should be less seen as a ratio of $\sigma(e^+e^-\to \mathcal{H})/\sigma(e^+e^-\to \mu^+\mu^-)$ but more simply as $\Gamma_{Z_Q \to \mathcal{H}}/\Gamma_{Z_Q\to \mu^+\mu^-}$.

\bibliographystyle{utphys}
\bibliography{ImprovedLVP}  

\end{document}